\documentclass[11pt,preprint]{aastex}

\newcommand{\syone}{\mbox{Sy 1}}
\newcommand{\sytwo}{\mbox{Sy 2}}
\newcommand{\syones}{\mbox{Sy 1's}}
\newcommand{\sytwos}{\mbox{Sy 2's}}
\newcommand{\ones}{\mbox{1's}}
\newcommand{\twos}{\mbox{2's}}
\newcommand{\kms}{\mbox{km\,s$^{-1}$}}
\newcommand{\whz}{\mbox{W\,Hz$^{-1}$}}
\newcommand{\cms}{\mbox{cm$^{-2}$}}
\newcommand{\etal}{et al.\ }
\newcommand{\oiii}{\mbox{[\ion{O}{3}]\,$\lambda$5007}}
\newcommand{\oiv}{\mbox{[\ion{O}{4}]\,$\lambda$25.9~\micron}}

\shorttitle{Spitzer IRS spectra of Seyfert galaxies}
\shortauthors{Buchanan et al.}

\begin{document}

\title{Spitzer IRS spectra of a large sample of Seyfert galaxies: a variety of
  infrared SEDs in the local AGN population}

\author{Catherine L.\ Buchanan\altaffilmark{1}, Jack F.\
Gallimore\altaffilmark{2}, Christopher P.\ O'Dea\altaffilmark{1}, Stefi A.\
Baum\altaffilmark{3}, David J.\ Axon\altaffilmark{1}, Andrew
Robinson\altaffilmark{1}, Moshe Elitzur\altaffilmark{4}, \& Martin
Elvis\altaffilmark{5}}

\altaffiltext{1}{Department of Physics, Rochester Institute of
Technology, 54 Lomb Memorial Drive, Rochester NY 14623. Email:
clbsps@cis.rit.edu}
\altaffiltext{2}{Bucknell University Department of Physics, Moore Avenue,
Lewisburg, PA 17837}
\altaffiltext{3}{Center for Imaging Science, Rochester Institute of
Technology, 84 Lomb Memorial Drive, Rochester NY 14623. Email:
clbsps@cis.rit.edu}
\altaffiltext{4}{University of Kentucky Physics \& Astronomy Department,
Lexington, KY 40506}
\altaffiltext{5}{Harvard-Smithsonian Center for Astrophysics, 60 Garden St.,
Cambridge, MA 02138}

\begin{abstract}
We are conducting a large observing program with the Spitzer Space Telescope
to determine the mid-to-far infrared spectral energy distributions of a
well-defined sample of 87 nearby, 12~\micron-selected Seyfert galaxies. In
this paper we present the results of IRS low-resolution spectroscopy of a
statistically representative subsample of 51 of the galaxies (59\%), with an
analysis of the continuum shapes and a comparison of the Seyfert types.  We
find that the spectra clearly divide into groups based on their continuum
shapes and spectral features.  The largest group (47\% of the sample of 51)
shows very red continuum suggestive of cool dust and strong emission features
attributed to PAHs. Sixteen objects (31\%) have a power-law continuum with
spectral indices $\alpha_{5-20\,\micron} = $-2.3 -- -0.9 that flatten to
$\alpha_{20-35\,\micron} = $-1.1 -- 0.0 at $\sim$20~\micron.  Clear silicate
emission features at 10 and 18~\micron\ are found in two of these objects
(Mrk~6 and Mrk~335).  A further 16\% of the sample show power-law continua
with unchanging slopes of $\alpha_{5-35\,\micron} = $-1.7 -- -1.1. Two objects
are dominated by a broad silicate absorption feature. One object in the sample
shows an unusual spectrum dominated by emission features, that is unlike any
of the other spectra.  Some spectral features are clearly related to a
starburst contribution to the IR spectrum, while the mechanisms producing
observed power-law continuum shapes, attributed to an AGN component, may be
dust or non-thermal emission.  The infrared spectral types appear to be
related to the Seyfert types.  Principal component analysis results suggest
that the relative contribution of starburst emission may be the dominant cause
of variance in the observed spectra. The derived starburst component of each
spectrum, however, contributes $<$40\% of the total flux density. We compare
the IR emission with the optically thin radio emission associated with the AGN
and find that \syones\ have higher ratios of IR/radio emission than \sytwos,
as predicted by the unified model if the torus is optically thick in the
mid-IR.  However, smooth-density torus models predict a much larger difference
between type \ones\ and \twos\ than the factor of 2 difference observed in our
sample; the observed factor of $\sim$2 difference between the type \ones\ and
\twos\ in their IR/radio ratios above 15~\micron\ requires the standard
smooth-density torus models to be optically thin at these wavelengths.
However, the resulting low torus opacity requires that the high observed
columns detected in X-ray absorption be produced in gas with very low dust to
gas ratio (perhaps within the dust sublimation region). On the other hand, our
observations may be consistent with clumpy torus models containing a steep
radial distribution of optically thick dense clumps.  The selection of our
sample at 12~\micron, where the torus may be optically thick, implies that
there may be orientation-dependent biases in the sample, however we do not
find that the sample is biased towards \sytwos\ with more luminous central
engines as would be expected.  We find that the \sytwos\ typically show
stronger starburst contributions than the \syones\ in the sample, contrary to
what is expected based on the unified scheme for AGN.  This may be due to the
selection effect that only those Seyfert \twos\ with strong starburst
contributions had high enough integrated 12~\micron\ flux densities to fall
above the flux limit of the sample.
\end{abstract}


\section{Introduction} \label{sec:intro}

Active galactic nuclei (AGNs) are powered by accretion onto a massive black
hole (see reviews by \citealt{ree84} and \citealt{shi99}). The fuel itself is
probably ISM from the host galaxy that is driven to the center by bar-induced
torques \citep{jog05}, galaxy interactions, or galaxy mergers (\citealt{ost93}
and references therein).  Starbursts may be an inevitable consequence of the
gas infall \citep{ree84,nor88}, and the observed relation between black hole
mass and host bulge mass in AGNs suggests that nuclear activity is closely
connected to star formation \citep{mer01,geb00}. Nuclear activity, though
probably short-lived compared with the lifetime of the host ($\sim$5 -- 20\%
percent of galaxies host AGN; \citealt{mai95a, ho97, kau03, hao05a}), may be
an integral part of the evolution of all galaxies. Understanding the inner
workings of AGNs and the AGN/starburst connection is therefore important for
understanding galaxy evolution.

Seyfert galaxies are the nearest and brightest AGNs and so are well-suited for
studying the connection between nuclear activity and star formation. Obscuring
dust however hampers studies of the optical -- soft X-ray emission produced by
young stars and the accretion disk. Emission at infrared wavelengths does not
suffer such large extinction, and, further, the dust that absorbs the shorter
wavelength emission reradiates in the IR and correspondingly produces a
substantial fraction of the bolometric flux of the object. Dissecting the
detailed IR spectral energy distribution (SED) of Seyfert galaxies can reveal
the properties of the dust in the nuclear region as well as place constraints
on the optical -- soft X-ray spectrum (i.e., starburst and AGN spectrum) that
heats the dust.

A key factor in understanding Seyfert galaxies is determining the geometry of
the nuclear obscuring material. Unification schemes postulate that the
observed differences between Seyfert types 1 and 2, such as optical emission
line widths and X-ray spectral slopes, are due to orientation-dependent
obscuration (see reviews by \citealt{urr95,ant93}).  Alternatively, there may
be fundamental differences between the central engines of \syones\ and \twos.
The obscuring medium has been modeled as a dusty torus which, viewed edge-on
(\sytwos), obscures the nucleus but, viewed face-on (\syones), allows the
nucleus to be seen.  The inner scale of the torus is thought to be of order
1~pc (e.g., \citealt{kro88, gal99, ulv99, jaf04, rot05}), but other properties
are poorly constrained owing to angular resolution limitations. However, the
geometry and equatorial opacity of the torus should have an observable impact
on the shape of the infrared SED \citep{pie92, gra94, ive97, nen02}. Thus AGN
unification can be tested by comparing the SEDs of type 1 and 2 Seyferts with
model predictions.

A third possibility is that the Seyfert types are connected by evolution: an
active nucleus evolves through a starburst (type 2) phase during which it is
heavily obscured, and into an AGN phase (type 1) when the obscuring dust has
been evacuated from the nucleus by starburst-driven winds and AGN outflows
\citep{san88, sto01}.  Supporting this picture is evidence that Seyfert \twos\
have significantly more star formation than Seyfert \ones. Photometric studies
have found that \sytwos\ have fainter nuclear infrared emission than \syones\
and SEDs consistent with starburst-dominated IR emission, while \syones\ have
SEDs consistent with pure AGN + quiescent galaxy emission \citep{ede87,
mai95b}.  However, as orientation is also likely to play a role in the
differences between Seyfert 1s and 2s, it is important to investigate the
relative contributions of star formation and the active nucleus to the
infrared emission in order to test both AGN unification and evolutionary
schemes.

The Spitzer space telescope represents a significant step forward in both
sensitivity and spatial resolution compared with previous thermal infrared
missions.  We are conducting a large observing program with the Spitzer space
telescope to determine the mid-to-far infrared (MFIR) spectral energy
distributions (SEDs) of 87 Seyfert galaxies (PI: J. Gallimore, PID: 3269). The
majority of the IRS spectra have been observed (51 objects, 59\%), allowing
the first statistical analysis of the detailed mid-infrared (MIR) SEDs of a
relatively unbiased sample of Seyfert galaxies. In this paper we present an
overview of the continuum properties of the sample, and relate these to the
optical spectral types. An atlas of spectra and detailed modeling of
individual spectra will be presented in a future paper when all data are
available. The purpose of this paper is to present the results of our initial
analysis of the spectra in order for them to be useful to the AGN community.
In \S\ref{sec:sample} we briefly describe the sample properties. The
observations and the data reduction are presented in
\S\ref{sec:obs}. \S\ref{sec:res} presents typical spectra and an analysis of
the continuum shapes.  In \S\ref{sec:dis} we discuss the implications of our
results for the unification of Seyfert galaxies and our understanding of the
nature of the torus.

\section{The Seyfert sample and its properties}\label{sec:sample}

The sample of Seyfert galaxies used for this study comprises all Seyfert
galaxies from the extended 12~\micron\ sample of \citet{rus93} that have $cz <
10000$~\kms. The potential advantage of the 12~\micron\ sample is its
perceived resistance to wavelength-dependent selection effects
\citep{sm89}. The redshift criterion was included to ensure that the smallest
possible physical size was probed by the fixed Spitzer apertures in order to
better exclude host galaxy emission. The resulting sample contains 87
sources. Three of the sample galaxies, NGC\,1097, NGC\,1566, and NGC\,5033,
are being extensively observed as part of the SINGS Legacy project, so we are
obtaining data for 84 galaxies.  The data include IRAC imaging, IRS spectral
mapping, and MIPS SED spectra.  The present paper discusses the 51 objects in
the sample for which 5 -- 35~\micron\ IRS spectra are currently available.

The spectral classifications were obtained from the \citet{ver03} catalog, and
thus have not been made using homogeneous datasets of consistent quality. In
several cases, the classification is ambiguous, and we have adopted revised
classifications from other sources \citep{tra03, ho97, vei95, phi79}.  Seven
objects in the sample have been reclassified as non-Seyferts (LINERs or
star-forming galaxies): MCG\,$+0-29-23$, NGC\,4922, NGC\,5953, NGC\,7130,
NGC\,7496, NGC\,7590, and UGC\,11680.  These objects have been excluded from
analyses comparing \syones\ and \sytwos, and we expect the results will not be
sensitive to the classifications of these objects. For the purposes of
comparing the Seyfert types we consider Seyfert types 1.$n$ (i.e. types 1.0,
1.2, 1.5, 1.8, and 1.9 as well as narrow line Seyfert 1's) to be \syones, and
objects with no broad permitted lines or with hidden broad line regions
(HBLRs) to be \sytwos.

\begin{figure}
\epsscale{0.3}
\plotone{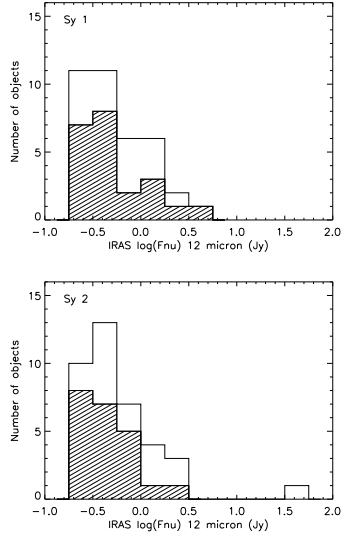}
\caption{Distribution of the IRAS 12~\micron\ fluxes for the Seyfert 1's {\it
  (top)} and the Seyfert 2's {\it (bottom)}. Hashed regions indicate the
  objects with IRS data in hand. Two-sample statistical tests show no
  significant difference between the \protect\syones\ and \protect\sytwos\ or
  between the observed objects and the whole sample (Table \ref{tab:statsa}).
  \label{fig:irashist}} \epsscale{1.0}
\end{figure}

\begin{figure}
\epsscale{0.3}
\plotone{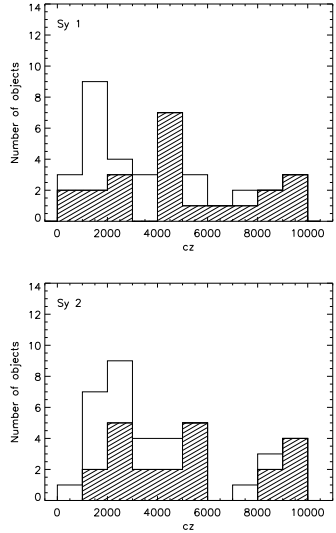}
\caption{ Distribution of redshifts for the 12~micron sample. Seyfert 1's {\it
  (top)} and Seyfert 2's {\it (bottom)} are shown separately. Hashed regions
  indicate the objects with IRS data in hand. Two-sample statistical tests
  show no significant difference between the \protect\syones\ and
  \protect\sytwos\ or between the observed objects and the whole sample (Table
  \ref{tab:statsa}).
  \label{fig:zhist}} \epsscale{1.0}
\end{figure}

Figures \ref{fig:irashist} and \ref{fig:zhist} show the distributions of the
IRAS 12~\micron\ flux densities and redshifts of the sample objects.  Data for
the whole sample are shown, with the objects observed to date with Spitzer
($\sim$ 50\% of the sample) indicated by hatching for comparison.  \syones\
and \sytwos\ are shown separately.  Two-sample statistical tests were
performed to determine whether the distributions of flux density and redshift
are similar for the Seyfert types and if the objects observed to date are
representative of the 12~\micron\ sample as a whole.  The probabilities of the
various datasets being drawn from the same parent population were calculated
using the ASURV package \citep{lav92}, which enables the analysis of data in
which limits are present \citep{fei85}.  The results of Gehan and logrank
tests are shown in Table \ref{tab:statsa}.  We find no significant differences
between \syones\ and \sytwos\ in either the IRAS 12~\micron\ flux density or
redshift distributions.  We also find no significant differences between the
observed subsample and the whole sample, indicating that the current dataset
is broadly representative of the complete sample.


\begin{figure}
\epsscale{0.5}
\plotone{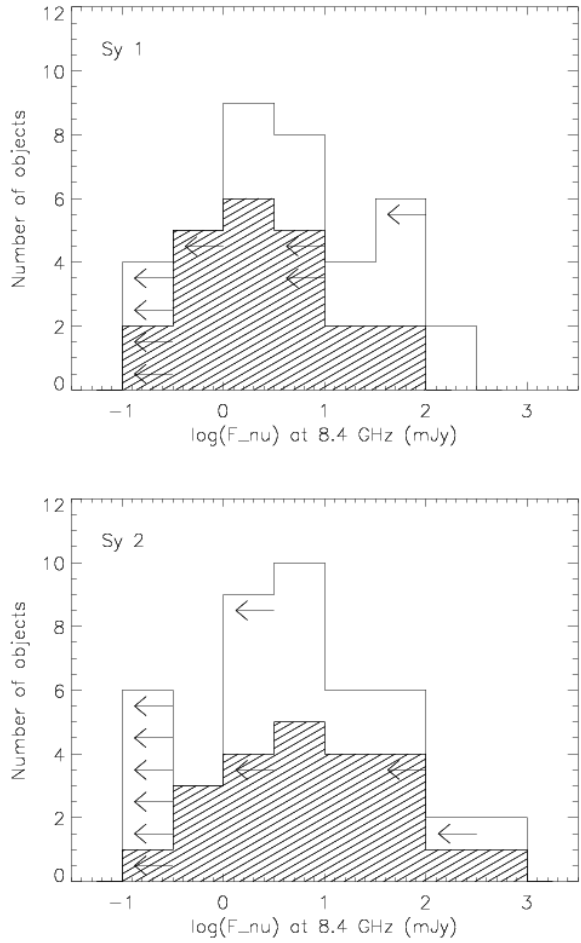}
\caption{Distribution of nuclear radio flux densities for the 12~micron\
  sample. Seyfert 1's {\it (top)} and Seyfert 2's {\it (bottom)} are shown
  separately. Hashed regions indicate the objects with IRS data in hand and
  arrows indicate upper limits. The results of two-sample tests comparing
  distributions are shown in Table \ref{tab:statsa}. \label{fig:radhist}}
  \epsscale{1.0}
\end{figure}

The optically thin nuclear radio emission may be considered an isotropic
property of Seyfert galaxies, that provides an orientation-independent
measurement of the energy output of the active nucleus
\citep{xu99,ede87b,giu90,kuk95,the01}. Radio observations of the 12~\micron\
Seyfert sample have been obtained by \citet{the00}, using the VLA A array at
8.4~GHz. These high spatial resolution (0.25\arcsec), high frequency
observations isolate the nuclear radio emission ($<$ 3.5\arcsec\ in extent)
associated with the AGN from radio emission associated with star formation in
the host galaxy. Seyfert nuclei are relatively weak radio emitters, so the
larger scale radio emission associated with star formation can dominate over
the nuclear radio emission at lower spatial resolutions.  Figure
\ref{fig:radhist} shows the distributions of nuclear radio flux densities for
the whole sample and for the objects for which we have IRS data in
hand. \syones\ and \sytwos\ are shown separately.  Limits are shown for 4
objects not detected by \citet{the00}. Radio flux densities for 6 objects not
observed by \citet{the00} were obtained from NED. These were generally lower
resolution and lower frequency, and so upper limits to the flux density at
8.4~GHz were derived assuming $S_{\nu} \propto \nu^{-0.7}$.  We find no
significant difference between the distributions of nuclear radio flux density
for the type 1 and 2 Seyferts.  Results of statistical comparisons of sample
properties are summarized in Table \ref{tab:statsa}.

\section{Observations and Spectral Extraction}\label{sec:obs}

The sample galaxies were observed by Spitzer IRS \citep{hou04} in mapping
mode, using the Short-Low and Long-Low modules. The modules cover a wavelength
range of $\sim$5 -- 35~\micron, with a resolving power of 64 -- 128. The raw
data were processed by the Spitzer pipeline version 11.0. The nuclear spectra
were extracted from the Basic Calibrated Data (BCDs) using SMART
\citep{hig04}.  Preliminary spectral maps show that the surface brightness is
dominated by a central point source at these wavelengths, so the default point
source aperture was used for the extractions. Only three of the sources
(NGC~1365, NGC~5005, \& NGC~5953) show a noticeable drop in flux between the
overlapping LL (long-low) and SL (short-low) module spectra, which results
from extended flux contributing to the wider LL module slit (SL: 3.5\arcsec,
LL: 10.5\arcsec). This supports our conclusion that the objects are dominated
by the central point source, and so default point source extractions are
appropriate. Flux densities were calibrated based on flux calibration tables
provided by the Spitzer Science Center. The spectra are shown in Figure
\ref{fig:allspec}, grouped according to their shape (see
\S\ref{sec:res_spe}). The full spectral maps will be presented in a future
paper.

Following a suggestion from the referee, comparison of the flux densities of
the nuclear IRS spectra with IRAC 5.8 and 8.0~\micron\ nuclear photometry
(Gallimore et al. 2005, in preparation) later revealed that the spectra
underestimated the source flux densities. We attribute this to slit losses
that vary as a function of wavelength due to the increasing size of the point
spread function with increasing wavelength.  As these spectra were obtained in
mapping mode, no attempt was made to center the nuclear point source in the
slit during the observations. Thus the nuclear (central) slit position is not
necessarily well centered on the source and the spectra suffer
wavelength-dependent slit losses, up to 20\%. This is only a problem for
strongly point-source dominated sources. We have attempted to account for this
by producing a spectral image cube from the Basic Calibrated Data FITS files
of all the slit positions (after background subtraction using the off orders)
and iteratively determining an interpolated nuclear image based on the
fractional shift perpendicular to the slit that produced the most total flux
in the interpolated image. A simplex maximization technique was used to find
the optimum interpolated position, which defined the plane containing the
spectral image of the point source.  We refer to the spectra extracted from
the spectral image of the point source as ``pointing-corrected'' nuclear
spectra.  Flux densities derived from the resulting pointing-corrected spectra
agree well with our IRAC photometry, within 20\% for the vast majority of
objects and 30-45\% for a few objects. The IRS-derived flux densities both
overestimate and underestimate the IRAC 5.8 and 8.0~\micron\ point source flux
densities, indicating that the flux calibration of our IRS spectra is not
systematically underestimating the true flux densities.  Excluding the
comparison with IRAS (\S\ref{sec:iras}), in the remainder of the paper the
analysis was performed on the original (not pointing-corrected) spectra.
Despite the fact that the flux errors are wavelength-dependent, we find that
the overall shapes of the spectra are not changed significantly by the
pointing correction. We do not expect that the flux errors will have a
significant impact on the results.

\begin{deluxetable}{lrrcrlcllccc}
\tablecolumns{12}
\tabletypesize{\scriptsize}
\rotate
\tablewidth{535pt}
\tablecaption{Properties of the sample: results of two-sample tests. \label{tab:statsa}}
\tablehead{
\colhead{Variable} &
\colhead{Group 1} &
\colhead{Group 2} &
\multicolumn{3}{c}{Group 1} &
\multicolumn{3}{c}{Group 2} &
\colhead{$p$} &
\colhead{$p$} &
\colhead{Figure}
 \\
 & & & 
\colhead{$n_{lim}$} &
\colhead{Mean} &
\colhead{Median} &
\colhead{$n_{lim}$} &
\colhead{Mean} &
\colhead{Median} &
\colhead{(Gehan)} &
\colhead{(Logrank)} &
 \\
\colhead{[1]} &
\colhead{[2]} & 
\colhead{[3]} & 
\colhead{[4]} &
\colhead{[5]} & 
\colhead{[6]} &
\colhead{[7]} &
\colhead{[8]} &
\colhead{[9]} &
\colhead{[10]} &
\colhead{[11]} &
\colhead{[12]} 
 \\
}
\startdata
IRAS $\log(F_{\rm 12~\micron})$  (Jy) & all \syones\  & all \sytwos\  & 0  & -0.28  & -0.37  &  0  & -0.23  & -0.35  & 0.63  & 0.60  &  \ref{fig:irashist} \\
IRAS $\log(F_{\rm 12~\micron})$  (Jy) & obs \syones\  & obs \sytwos\  & 0  & -0.31  & -0.39  &  0  & -0.34  & -0.42  & 0.99  & 0.62  &  \ref{fig:irashist} \\
IRAS $\log(F_{\rm 12~\micron})$  (Jy) & all \syones\  & obs \syones\  & 0  & -0.28  & -0.37  &  0  & -0.31  & -0.39  & 0.69  & 0.90  &  \ref{fig:irashist} \\
IRAS $\log(F_{\rm 12~\micron})$  (Jy) & all \sytwos\  & obs \sytwos\  & 0  & -0.23  & -0.35  &  0  & -0.34  & -0.42  & 0.30  & 0.26  &  \ref{fig:irashist} \\
$z$                                   & all \syones\  & all \sytwos\  & 0  & 0.0139 & 0.0133 &  0  & 0.0143 & 0.0117 & 0.80  & 0.99  &  \ref{fig:zhist} \\
$z$                                   & obs \syones\  & obs \sytwos\  & 0  & 0.0169 & 0.0162 &  0  & 0.0171 & 0.0172 & 0.83  & 0.99  &  \ref{fig:zhist} \\
$z$                                   & all \syones\  & obs \syones\  & 0  & 0.0139 & 0.0133 &  0  & 0.0169 & 0.0162 & 0.23  & 0.29  &  \ref{fig:zhist} \\
$z$                                   & all \sytwos\  & obs \sytwos\  & 0  & 0.0143 & 0.0117 &  0  & 0.0171 & 0.0172 & 0.22  & 0.28  &  \ref{fig:zhist} \\
$\log(S_{\rm 8.4~GHz})$ (mJy)         & all \syones\  & all \sytwos\  & 8  &  0.57* & 0.51   &  6  & 0.83*  & 0.88   & 0.20  & 0.29  &  \ref{fig:radhist} \\
$\log(S_{\rm 8.4~GHz})$ (mJy)         & obs \syones\  & obs \sytwos\  & 5  &  0.40* & 0.40   &  2  & 0.89*  & 1.05   & 0.05  & 0.11  &  \ref{fig:radhist} \\
$\log(S_{\rm 8.4~GHz})$ (mJy)         & all \syones\  & obs \syones\  & 8  &  0.57* & 0.51   &  5  & 0.40*  & 0.40   & 0.46  & 0.48  &  \ref{fig:radhist} \\
$\log(S_{\rm 8.4~GHz})$ (mJy)         & all \sytwos\  & obs \sytwos\  & 6  &  0.83* & 0.88   &  2  & 0.89*  & 1.05   & 0.73  & 0.73  &  \ref{fig:radhist} \\
\enddata 
\tablecomments{Generalized Wilcoxon and Logrank two-sample tests were used to
determine the probability that the values of the variable (column [1]) in
groups 1 and 2 (columns [2] and [3]) were drawn from the same parent
distribution.  The number of objects in the groups are: all \protect\syones:
37; observed \protect\syones: 22; all \protect\sytwos: 38; observed
\protect\sytwos: 22.  Columns [4]-[6] list the parameters for group 1: the
number of limits in the data, the mean value and the median value, derived
using the Kaplan-Meier (K-M) estimator. * Indicates the mean is biased because
the lower- or uppermost limit was changed to a detection to compute the K-M
distribution.  Columns [7]-[9] list the same parameters for group 2.  Large
values of $p$ (columns [10] and [11]) indicate there is no significant
difference between the distributions for the two groups. Column [12] refers to
the Figure number showing the distribution in the variables. }
\end{deluxetable}

\begin{figure}
\plottwo{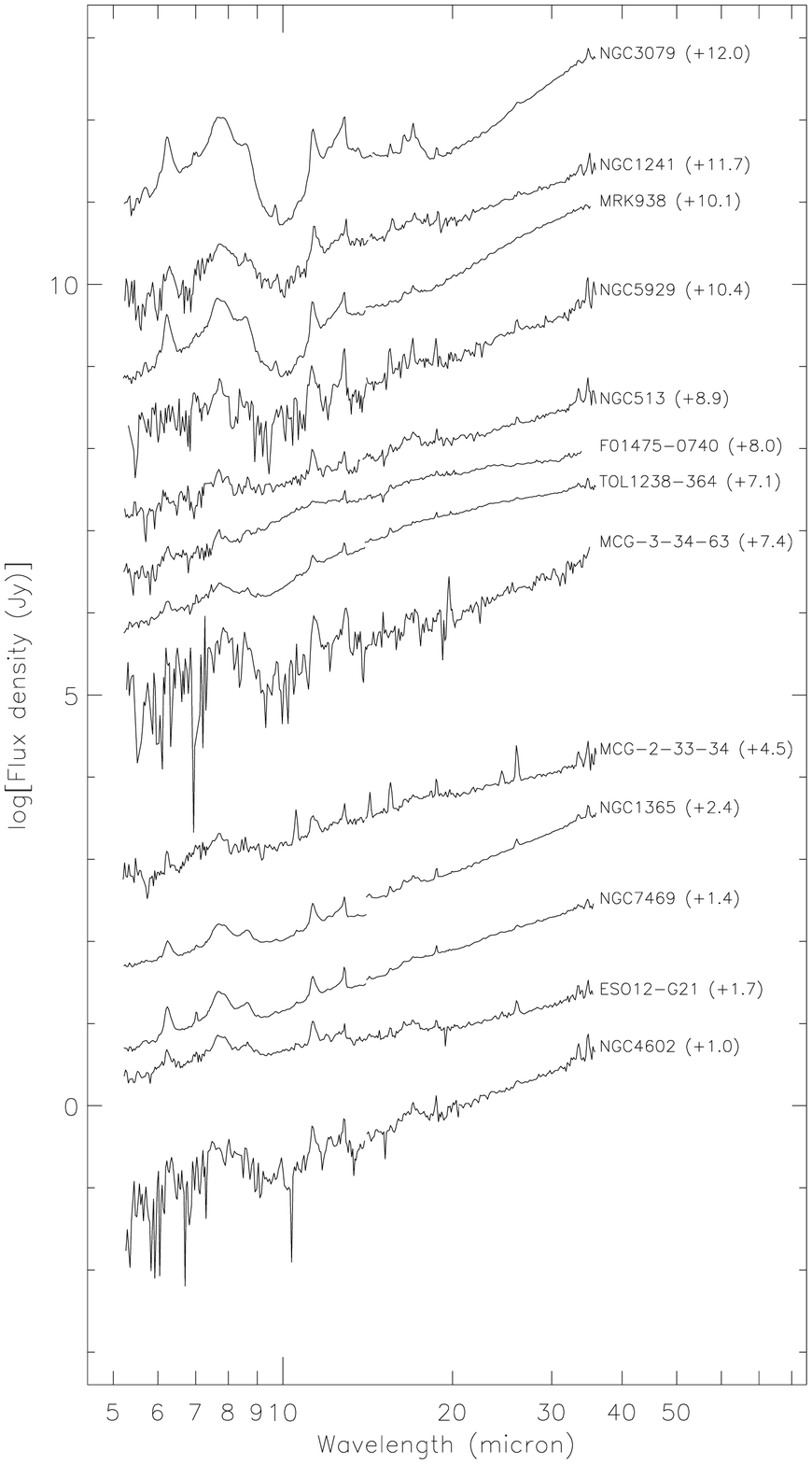}{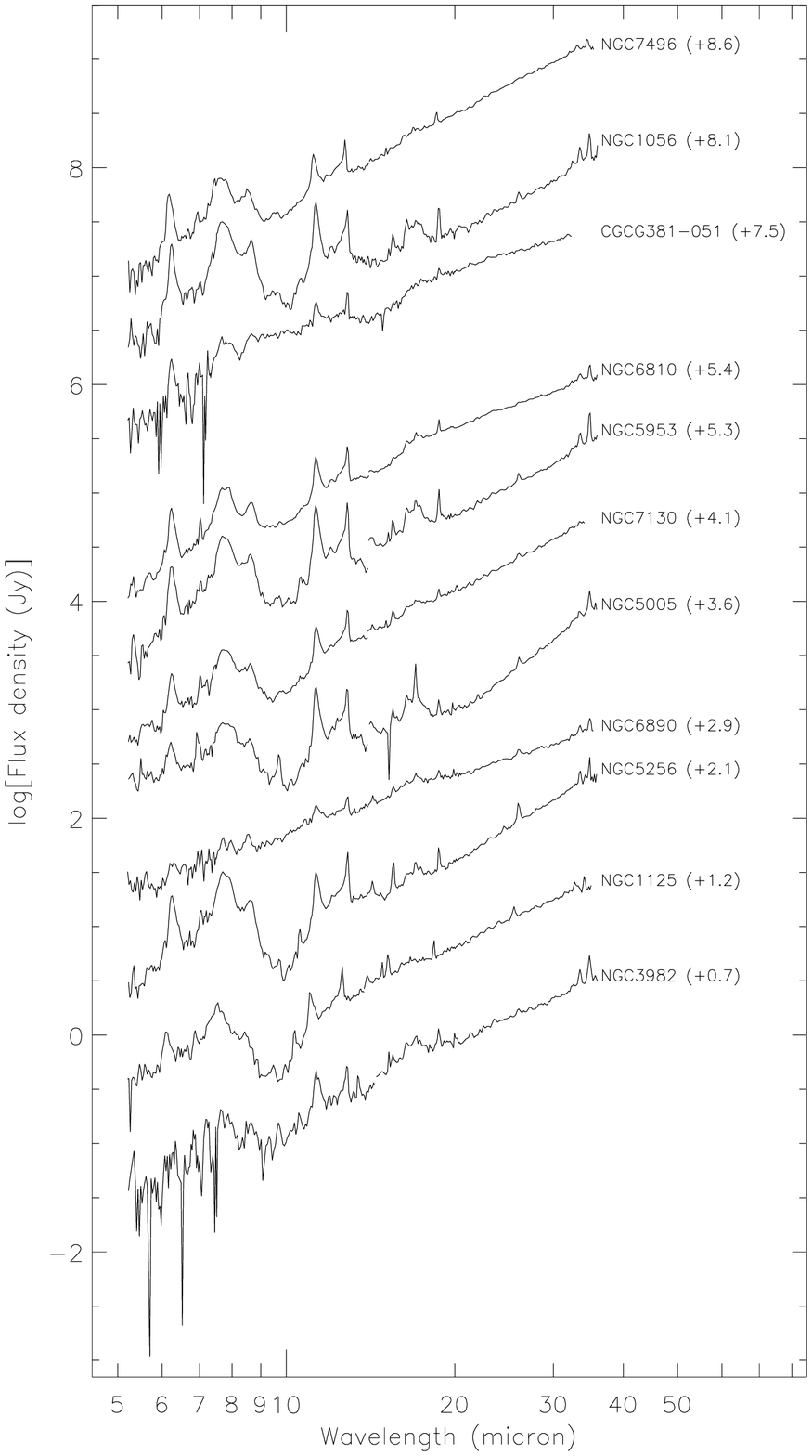}
\caption{{\it (a)} Spectra dominated by red continuum and PAH features. Each
  spectrum is labeled with the object name and the flux density offset
  applied.  \label{fig:allspec}}
 \epsscale{1.0}
\end{figure}

\clearpage

\begin{figure}
\addtocounter{figure}{-1}
\plottwo{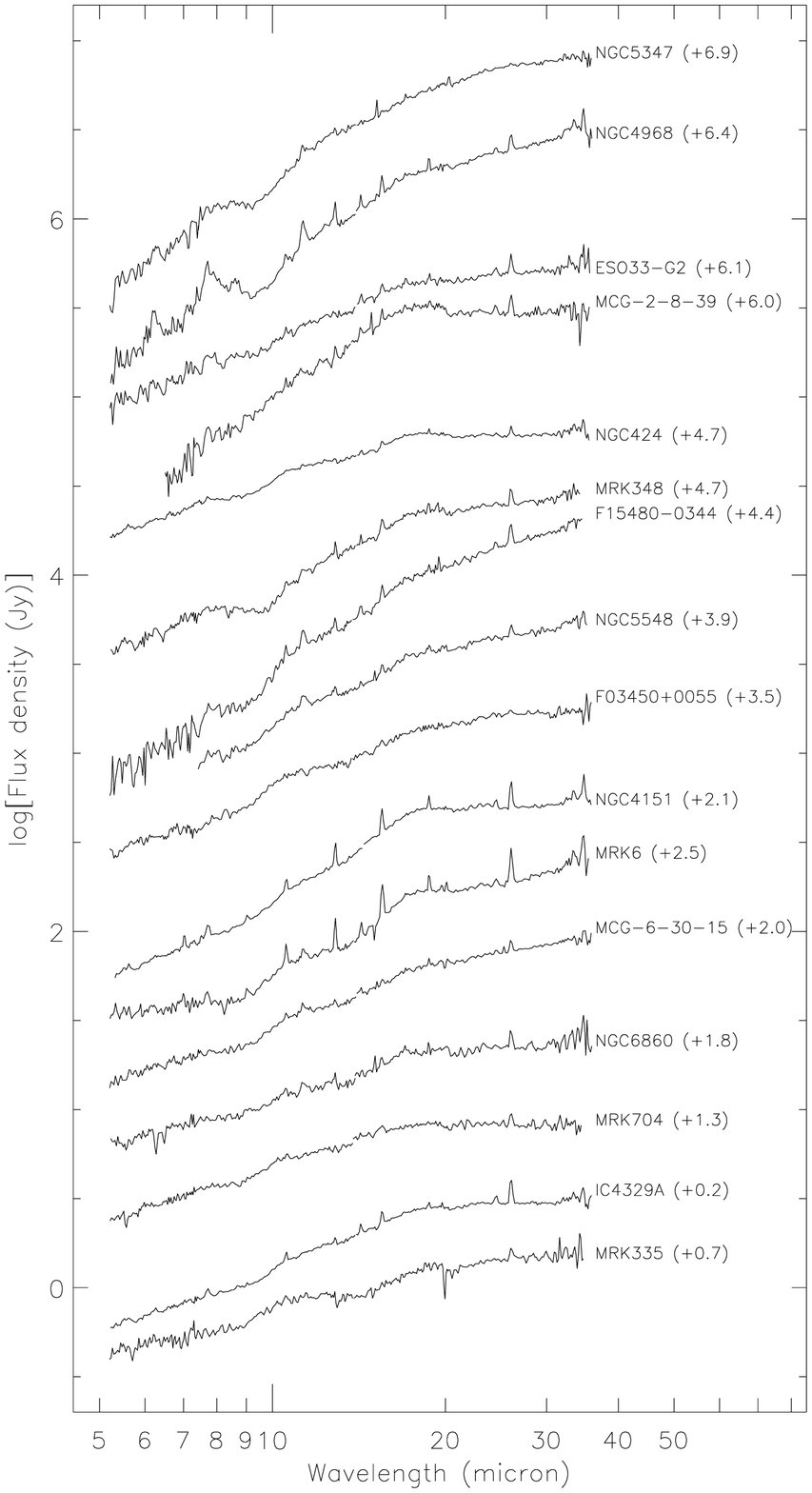}{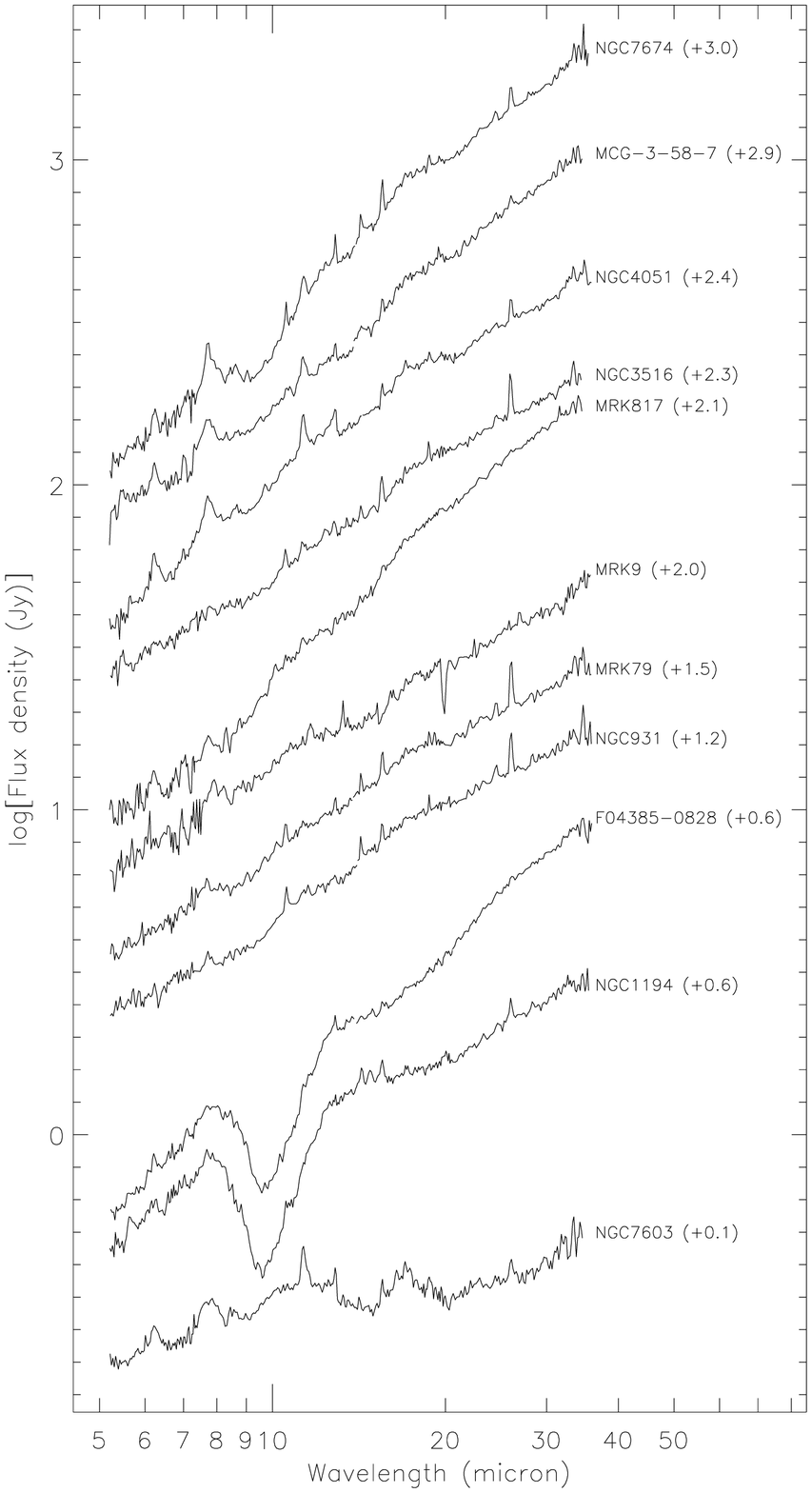}
\caption{{\it {(b)}} The left panel shows spectra that may be described by a
  broken power law continuum. The right panel shows spectra that can be fitted
  with a continuous power law, the two spectra showing silicate absorption at
  10~\micron, and the ungrouped spectrum (bottom). Each
  spectrum is labeled with the object name and the flux density offset
  applied.}  \epsscale{1.0}
\end{figure}

\clearpage

\section{Results} \label{sec:res}

\subsection{Comparison with IRAS}\label{sec:iras}

\begin{figure}
\epsscale{0.9}
\plottwo{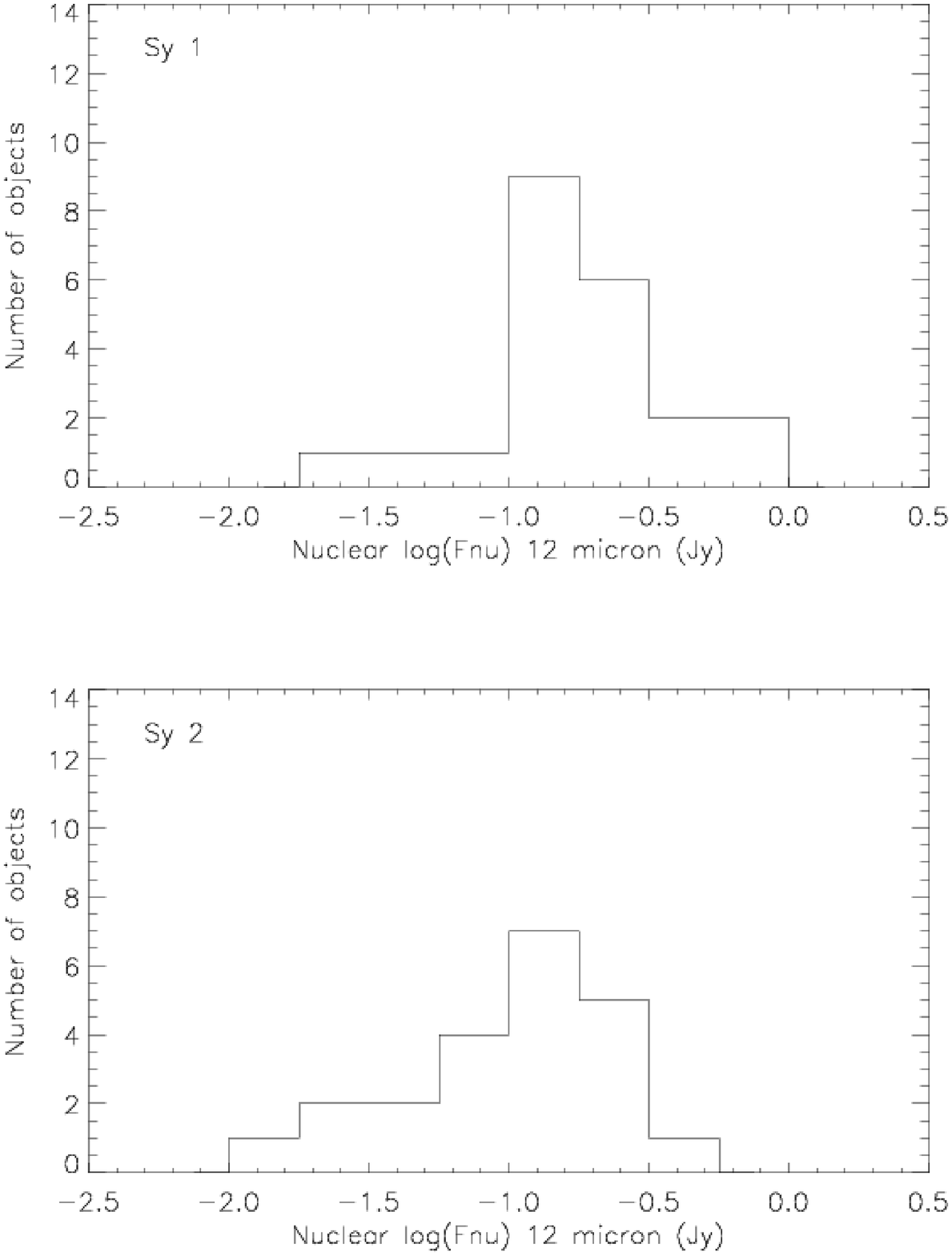}{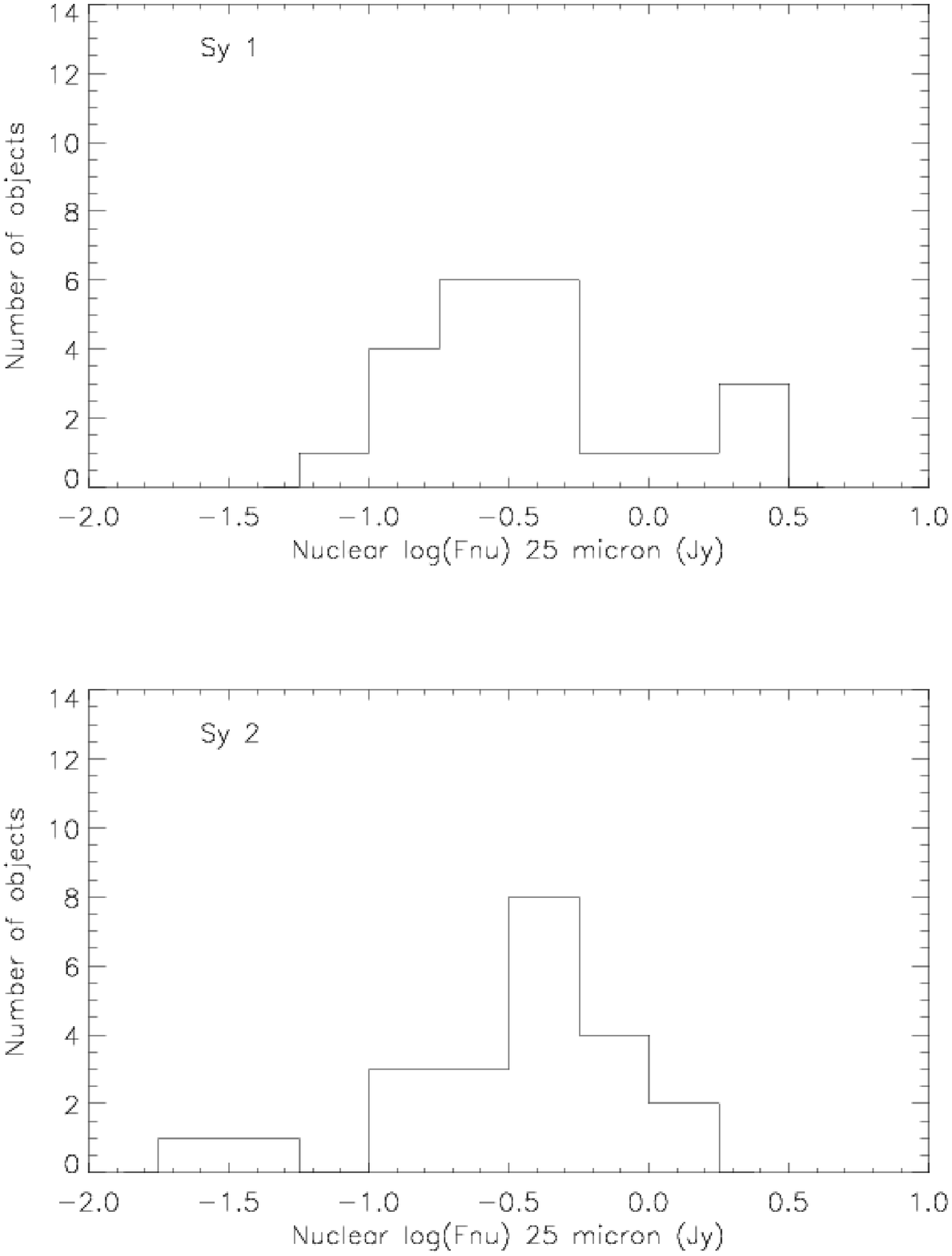}
\caption{Distribution of the 12~\micron\ {\it (left)} and 25~\micron {\it
  (right)} nuclear fluxes measured from the IRS spectra. Seyfert 1's {\it
  (top)} and Seyfert 2's {\it (bottom)} are shown separately.
  \label{fig:fluxhist}} 
\epsscale{1.0}
\end{figure}

The IRS apertures are much smaller than those of IRAS, and so we can evaluate
the fraction of extended flux not included in the IRS aperture (assuming
insignificant mid-infrared variation in the $\sim$ 24 years between
observations).  The IRS spectra span the wavelength range 5 -- 35~\micron, so
12 and 25~\micron\ nuclear flux densities corresponding to the IRAS bandpasses
can be derived from the nuclear spectra. For the purposes of this comparison,
the nuclear flux densities were determined using the pointing-corrected IRS
spectra and the \mbox{IRAS} bandpass spectral response functions. The
distributions of nuclear 12 and 25~\micron\ flux densities for \syones\ and
\sytwos\ are shown in Figure \ref{fig:fluxhist}.  There is a marginal
probability that the distributions for \syones\ and \sytwos\ are different at
12~\micron, with the \sytwos\ fainter than the \syones\ (7\% probability that
the difference in the observed distributions ocurrs due to chance; Table
\ref{tab:statsb}), but the distributions are statistically similar at
25~\micron.

\begin{figure}
\epsscale{0.9}
\plottwo{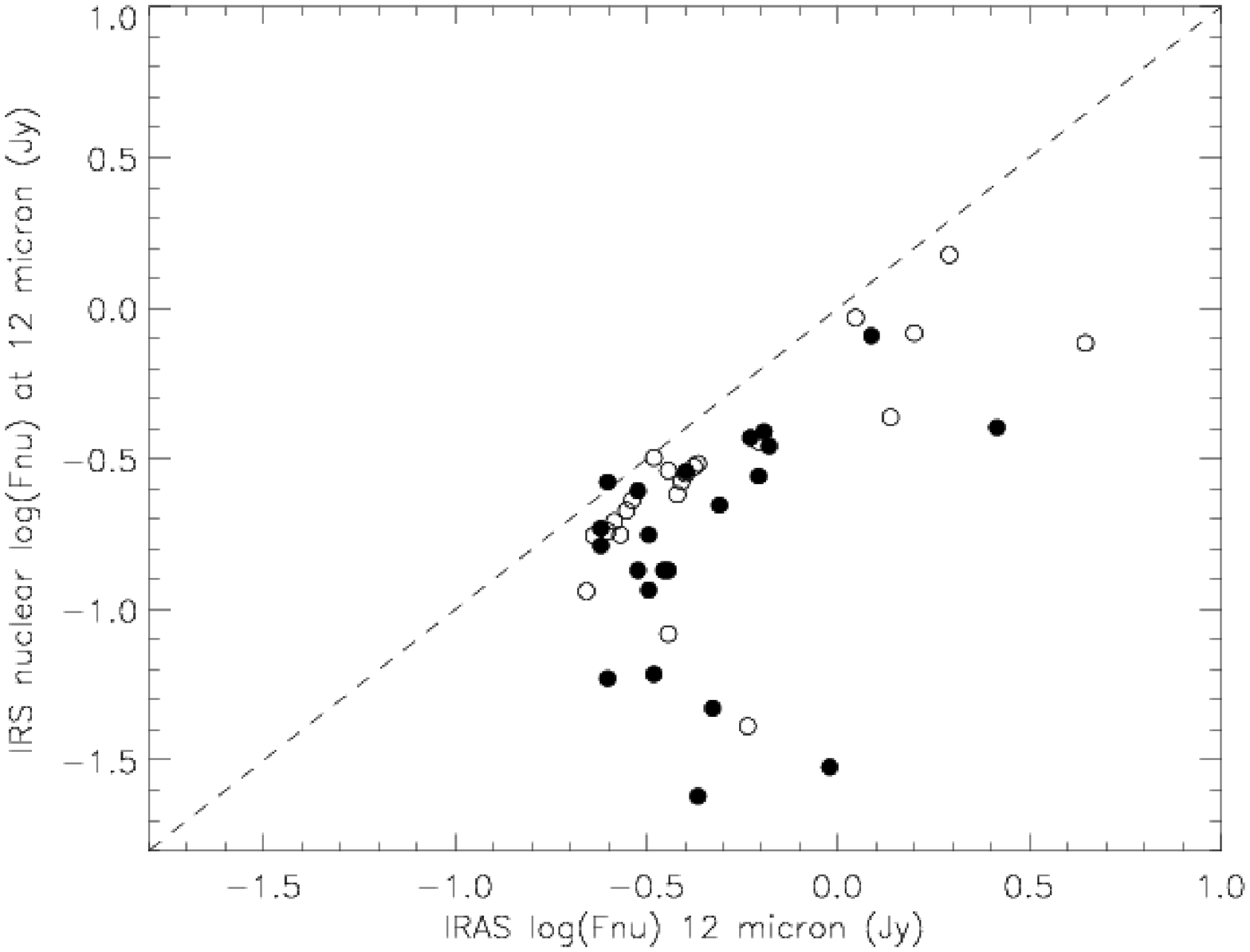}{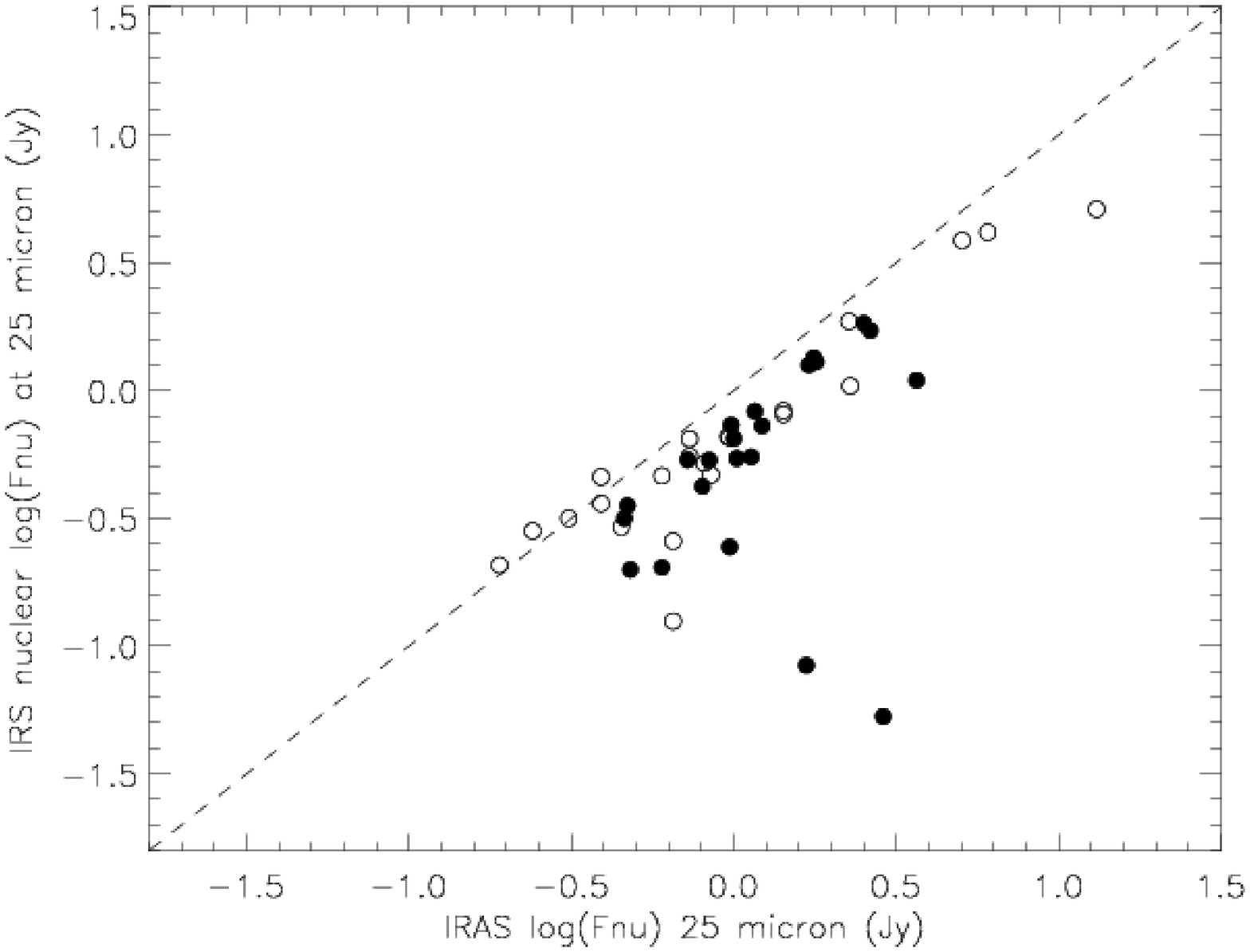}
\caption{IRAS global flux density vs IRS nuclear flux density at 12~\micron\
  {\it (left)} and 25~\micron {\it (right)}. Seyfert 1's {\it (open circles)}
  and Seyfert 2's {\it (filled circles)} are shown separately. The dashed
  lines indicate equal global and nuclear fluxes.
  \label{fig:fluxnucglob}}
\epsscale{1.0}
\end{figure}

\begin{figure}
\epsscale{0.8}
\plottwo{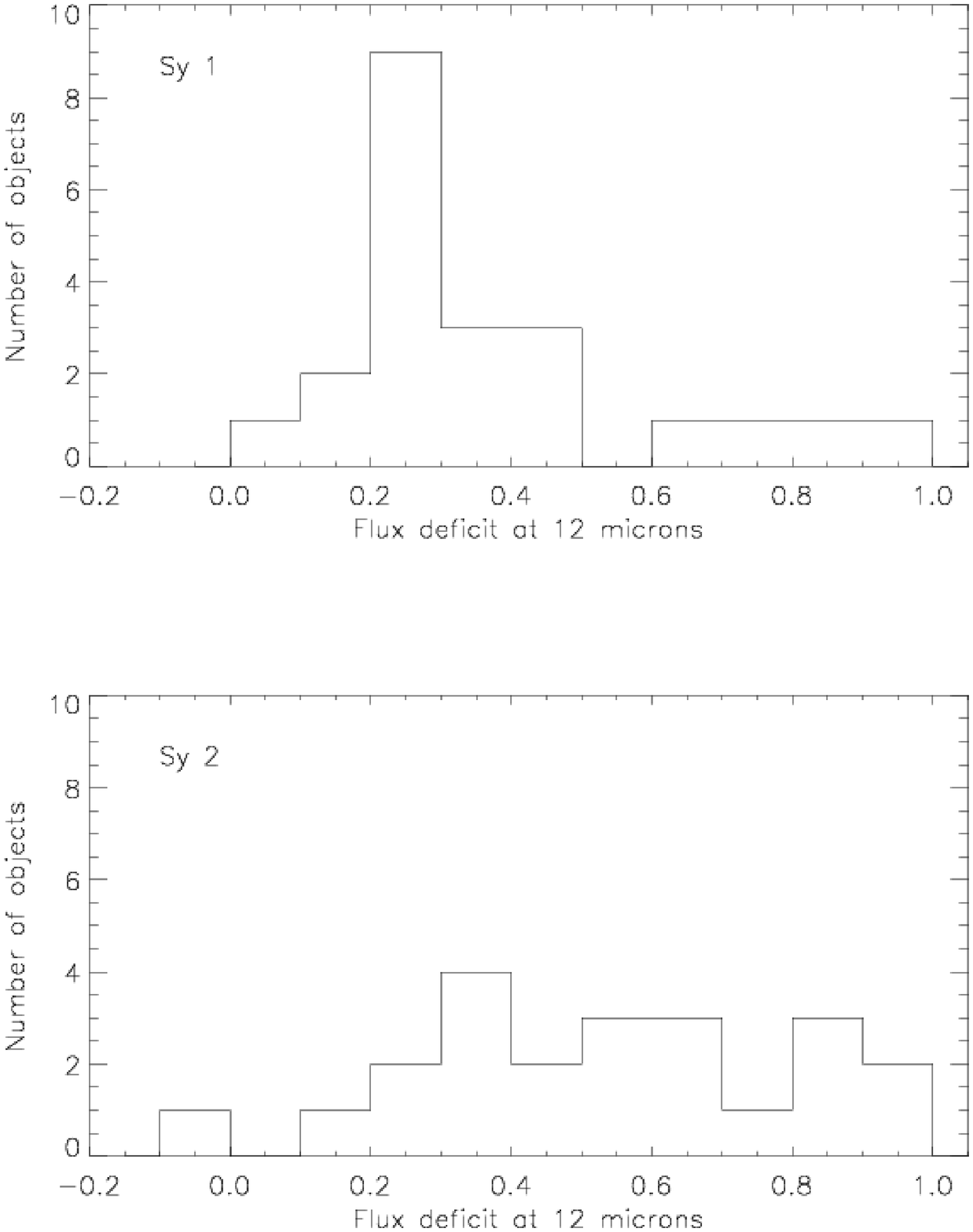}{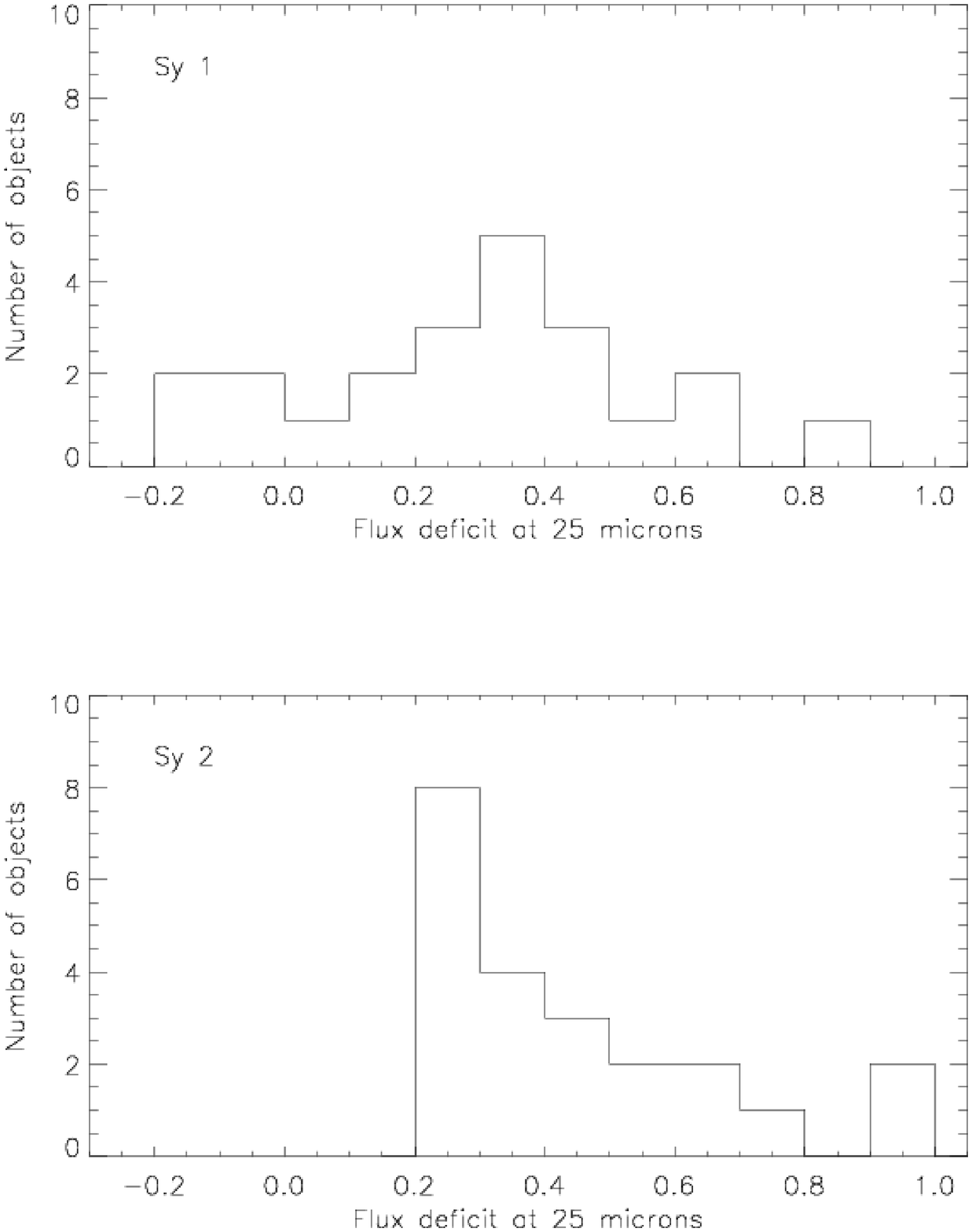}
\caption{Distribution of the flux deficit, the fraction of flux missing in the
  IRS spectra at 12~\micron\ {\it (left)} and 25~\micron {\it (right)}
  compared with the IRAS flux densities.
  \label{fig:fluxmiss}} 
\epsscale{1.0}
\end{figure}

We calculate the flux deficit, the fraction of flux ``missing'' in the IRS
aperture relative to IRAS, $F_{def} = (F_{IRAS}-F_{IRS})/F_{IRAS}$.  Figures
\ref{fig:fluxnucglob} and \ref{fig:fluxmiss} show that the sample includes
some objects for which all the IRAS flux is concentrated within the IRS
low-resolution slits (widths 3.5 -- 10.5\arcsec), and other objects for which
most of the flux falls outside the slit. For most objects, the majority of the
IRAS flux is seen by IRS, indicating the surface brightness distributions of
the objects are compact. For a few objects, the IRS flux density is higher
than the IRAS flux density, suggesting that variability may be contributing to
the scatter in the distributions of flux deficit; however, the differences are
$\lesssim$20\%, which is within our flux density uncertainties. These results
are consistent with the results of previous studies that compared ground-based
photometry of Seyfert galaxies with IRAS photometry. \citet{mai95b} compared
5\arcsec\ aperture 10~\micron\ flux densities of Seyfert galaxies with the
IRAS measurements and showed that the Seyferts displayed a range of
ground/IRAS flux densities from 0.0 -- 1.0, indicating that the sources ranged
from highly compact objects dominated by the nuclear point source, to those
with relatively strong extended emission, associated with star-formation.
\citet{gor04} also compared small aperture (1.5\arcsec) ground-based
photometry of Seyfert galaxies with IRAS measurements.  Their results show a
similar distribution of flux deficits to our dataset, and these authors note
that their data are relatively insensitive to diffuse extended emission due to
the high spatial resolution of the detector, and to the small chopping
distance used in the observations, which likely resulted in over-subtraction
of background emission.

We find that the average flux deficits of \sytwos\ (means of 0.71 and 0.66 at
12 and 25~\micron, respectively) are greater than those of \syones\ (means of
0.62 and 0.56; Figures \ref{fig:fluxnucglob} and \ref{fig:fluxmiss}), although
statistical comparisons indicate that the differences are only marginally
significant (formal probabilities that the parent distributions are the same
are 6 -- 11\%; Table \ref{tab:statsb}).  The lack of a jump in flux density
between the spectra obtained with the SL and LL modules, which have different
slit widths, suggests that the emission occurs on scales larger than the LL
module slit width ($\sim$10\arcsec).  This implies that much of the MIR
emission is on kiloparsec scales, and that this difference between the
\syones\ and \twos\ is related to relatively extended star formation rather
than orientation-dependent obscuration in the nucleus.  The difference in flux
deficits between the \syones\ and \sytwos\ implies that the \sytwos\ have {\it
relatively} more extended emission than the \syones, that is, they are less
dominated by the nuclear point source.  If both types of Seyfert have the same
luminosities, then our results would indicate that the type \twos\ have more
luminous extended emission than the type \ones.  As noted above, the
distributions of nuclear 25~\micron\ flux density for the \syones\ and
\sytwos\ are similar, while at 12~\micron\ the nuclear flux densities of type
\ones\ and \twos\ are marginally significantly different
(Fig. \ref{fig:fluxhist} and Table \ref{tab:statsb}).  We note that, as the
sample was selected at 12~\micron, \sytwos\ with weak nuclei may have been
excluded, resulting in the 12~\micron\ distributions of type 1's and 2's
appearing more similar than they in fact are; the effects of sample selection
on our results are discussed further in \S\ref{subsec:dis_sel}.  Therefore it
would appear that the \sytwos\ have more extended emission relative to the
nuclear emission than the \syones\ {\it and} that the \sytwos\ have, on
average, more luminous extended emission than the type \ones.


\begin{deluxetable}{lrrcrlcllccc}
\tablecolumns{12}
\tabletypesize{\scriptsize}
\rotate
\tablewidth{554pt}
\tablecaption{Properties derived from IRS spectra: results of two-sample tests. \label{tab:statsb}}
\tablehead{
\colhead{Variable} &
\colhead{Group 1} &
\colhead{Group 2} &
\multicolumn{3}{c}{Group 1} &
\multicolumn{3}{c}{Group 2} &
\colhead{$p$} &
\colhead{$p$} &
\colhead{Figure}
 \\
 & & & 
\colhead{$n_{lim}$} &
\colhead{Mean} &
\colhead{Median} &
\colhead{$n_{lim}$} &
\colhead{Mean} &
\colhead{Median} &
\colhead{(Gehan)} &
\colhead{(Logrank)} &
 \\
\colhead{[1]} &
\colhead{[2]} & 
\colhead{[3]} & 
\colhead{[4]} &
\colhead{[5]} & 
\colhead{[6]} &
\colhead{[7]} &
\colhead{[8]} &
\colhead{[9]} &
\colhead{[10]} &
\colhead{[11]} &
\colhead{[12]} 
 \\
}
\startdata
IRS $\log(F_{{\rm 12\,\micron}})$ (Jy) & obs \protect\syones\  & obs \protect\sytwos\  & 0  & -0.76  & -0.78  &  0  & -0.99  & -0.92  & 0.07  & 0.07  &  \ref{fig:fluxhist} \\
IRS $\log(F_{{\rm 25\,\micron}})$ (Jy) & obs \protect\syones\  & obs \protect\sytwos\  & 0  & -0.41  & -0.47  &  0  & -0.52  & -0.47  & 0.78  & 0.44  &  \ref{fig:fluxhist} \\
12~\micron\ flux deficit & obs \protect\syones\  & obs \protect\sytwos\  & 0  &  0.62  & 0.57   &  0  & 0.71   & 0.70   & 0.06  & 0.11  &  \ref{fig:fluxmiss} \\
25~\micron\ flux deficit & obs \protect\syones\  & obs \protect\sytwos\  & 0  &  0.56  & 0.57   &  0  & 0.66   & 0.63   & 0.06  & 0.07  &  \ref{fig:fluxmiss} \\
E/v 1 component & red, PAH  & broken P-L     & 0  & 14.1  & 14.1  &  0  & -7.3   &  -7.3  & 0.0000  & 0.0000  & \ref{fig:evonehist} \\
E/v 1 component & red, PAH  & unbroken P-L   & 0  & 14.1  & 14.1  &  0  & -1.1   &  -2.1  & 0.0001  & 0.0000  & \ref{fig:evonehist} \\
E/v 1 component & broken P-L  & unbroken P-L   & 0  & -7.3  & -7.3  &  0  & -1.1   &  -2.1  & 0.0003  & 0.0002  & \ref{fig:evonehist} \\
E/v 1 component & obs \protect\syones\ & obs \protect\sytwos\      & 0  & -1.2  & -2.9  &  0  & 6.2   &  3.3  & 0.04  & 0.02  & \ref{fig:evonesy} \\
$\log(F_{{\rm 10.6\,\micron}}/S_{{\rm 8.4\,GHz}})$   & obs \protect\syones\  & obs \protect\sytwos\  & 5  &  2.01*  & 1.91   &  2  & 1.16*   & 1.25    & 0.002  & 0.011  & \ref{fig:irradio} \\
E/v 1 component & high $L_{{\rm 8.4\,GHz}}$ & low $L_{{\rm 8.4\,GHz}}$ & 0  &  5.6  &   0.7  &  0  &  4.7   & 2.9  & 0.88  & 0.59  & \ref{fig:radpowerhist} \\
\enddata 
\tablecomments{Generalized Wilcoxon and Logrank two-sample tests were used to
determine the probability that the values of the variable (column [1]) in
groups 1 and 2 (columns [2] and [3]) were drawn from the same parent
distribution.  The number of objects in the groups are: all \protect\syones:
37; observed \protect\syones: 22; all \protect\sytwos: 38; observed
\protect\sytwos: 22.  Columns [4]-[6] list the parameters for group 1: the
number of limits in the data, the mean value and the median value, derived
using the Kaplan-Meier (K-M) estimator where there are limits in the sample. *
Indicates the mean is biased because the lower- or uppermost limit was changed
to a detection to compute the K-M distribution.  Columns [7]-[9] list the same
parameters for group 2.  Large values of $p$ (columns [10] and [11]) indicate
there is no significant difference between the distributions for the two
groups. Column [12] refers to the Figure number showing the distribution in
the variables. }
\end{deluxetable}

\subsection{Spectral shapes} \label{sec:res_spe}

We find four distinct types of continuum shapes and spectral features among
the Seyfert nuclear spectra.  Figure \ref{fig:egspec} shows typical spectra in
each group. An atlas of spectra will be presented when all data have been
obtained.  All spectra have been converted to the restframe. We group the
spectra according to the following main properties:\\
$\bullet$ Twenty-four objects (47\%\ of the observed objects) have
PAH-dominated spectra with strong emission features at 6.2, 7 -- 9, 11.3, and
12.9~\micron\ and very red continuum suggestive of cool dust; Mrk\,938 is the
archetype (Fig.~\ref{fig:egspec}). Eight objects, including Mrk\,938, show
clear silicate absorption at 10~\micron, while a further eight show possible
weak absorption, but the continuum level is difficult to determine due to the
strong PAH emission. The remaining 8 objects in this group show no apparent
silicate features.  There are 5 \syones, 12 \sytwos, 3 LINERs, and 4 starburst
galaxies in this group. \\
$\bullet$ Sixteen objects (31\%) have continua that can be described by a
broken power-law; they show a flattening in the continuum slope at
$\sim$20~\micron; NGC\,4151 is the brightest of this class
(Fig.~\ref{fig:egspec}). This flattening may be due to a warm ($\sim$170~K)
dust component, peaking at $\sim$20~\micron, dominating the mid-IR emission
(e.g., \citealt{wee05,per01,rod96}), however, a simple model of an
emissivity-weighted blackbody function does not fit the continuum of these
spectra well. Models including multiple dust components are required to
determine if this is responsible for the continuum shape (Buchanan et al., in
prep.).  Two of these objects, Mrk6 and Mrk335, show clear silicate emission
features (Figure \ref{fig:silspec}{\it b}). A further 9 objects in this group,
including NGC4151 (Fig. \ref{fig:egspec}), show weak excesses at 10 and
18~\micron\ that may be due to silicate emission; the features are
sufficiently weak that a more quantitative analysis is necessary to confirm
the presence of silicate dust emission in these spectra. One object, NGC5347,
shows apparent weak silicate absorption at 10~\micron\ in its spectrum, while
the remaining 4 objects have ambiguous spectral features.  There are 9
\syones, and 7 \sytwos\ in this group. Power-law fits to these spectra
indicate that the spectral indices below 20~\micron\ range from -2.3 to -0.9,
with a median value of $\alpha_{5-20}$~$=$~-1.3 (where $S_{\nu} \propto
\nu^{\alpha}$). The spectral indices above the break are -1.1 -- 0.0, with the
median $\alpha_{20-35}$~$=$~-0.4. \\
$\bullet$ Eight objects (16\%) have power-law continuous spectra over the IRS
spectral range $\sim$5 -- 35~\micron. NGC~3516 is representative of this class
(Fig.~\ref{fig:egspec}). These spectra show no strong dust emission or
absorption features, though most have weak excesses at 10 and 18~\micron\ that
may be due to silicate emission (Fig. \ref{fig:egspec}).  Six of these objects
are \syones, and two are \sytwos. Power-law fits to these data indicate the
spectral indices range from -1.7 to -1.1, with a median of
$\alpha_{5-35}$~$=$~-1.2. \\
$\bullet$ Two objects (NGC\,1194 \& F04385$-$0828) show a broad absorption
feature at 9.7~\micron\ due to silicate dust. NGC\,1194 is a \syone, and
F04385$-$0828 is a \sytwo. \\
$\bullet$ One object (NGC\,7603) appears to show both PAH and silicate
emission features and otherwise fails to meet any of the above
classifications. This object is a \syone.

\begin{figure}
\epsscale{0.8}
\plotone{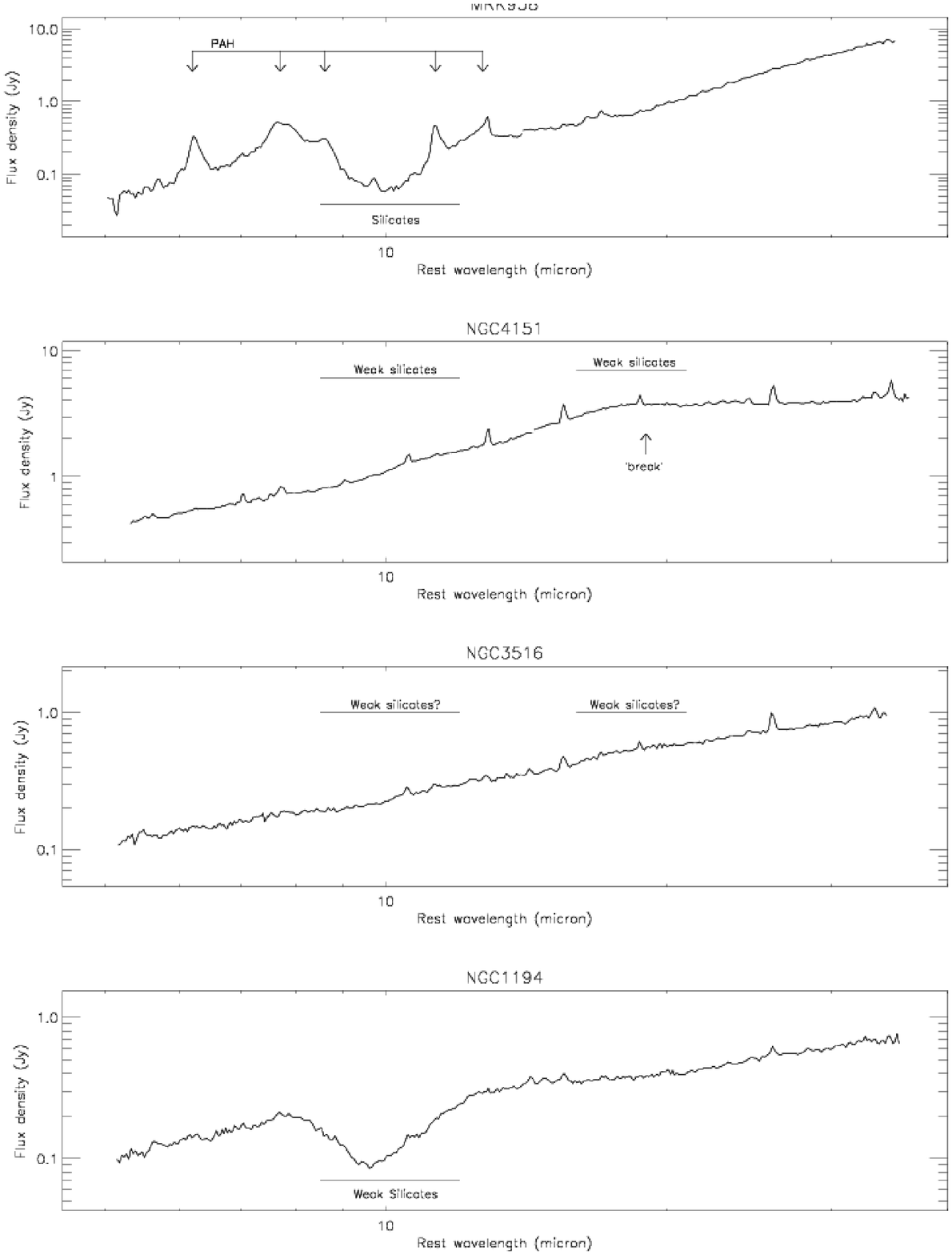}
\caption{{\it (a)} Typical spectra in each group: Mrk 938 (red continuum with
  PAH features), NGC4151 (broken power-law), NGC3516 (power-law), and NGC1194
  (silicate absorption). All spectra have been converted to restframe.
  \label{fig:egspec}} \epsscale{1.0}
\end{figure}

\begin{figure}
\addtocounter{figure}{-1}
\epsscale{0.8}
\plotone{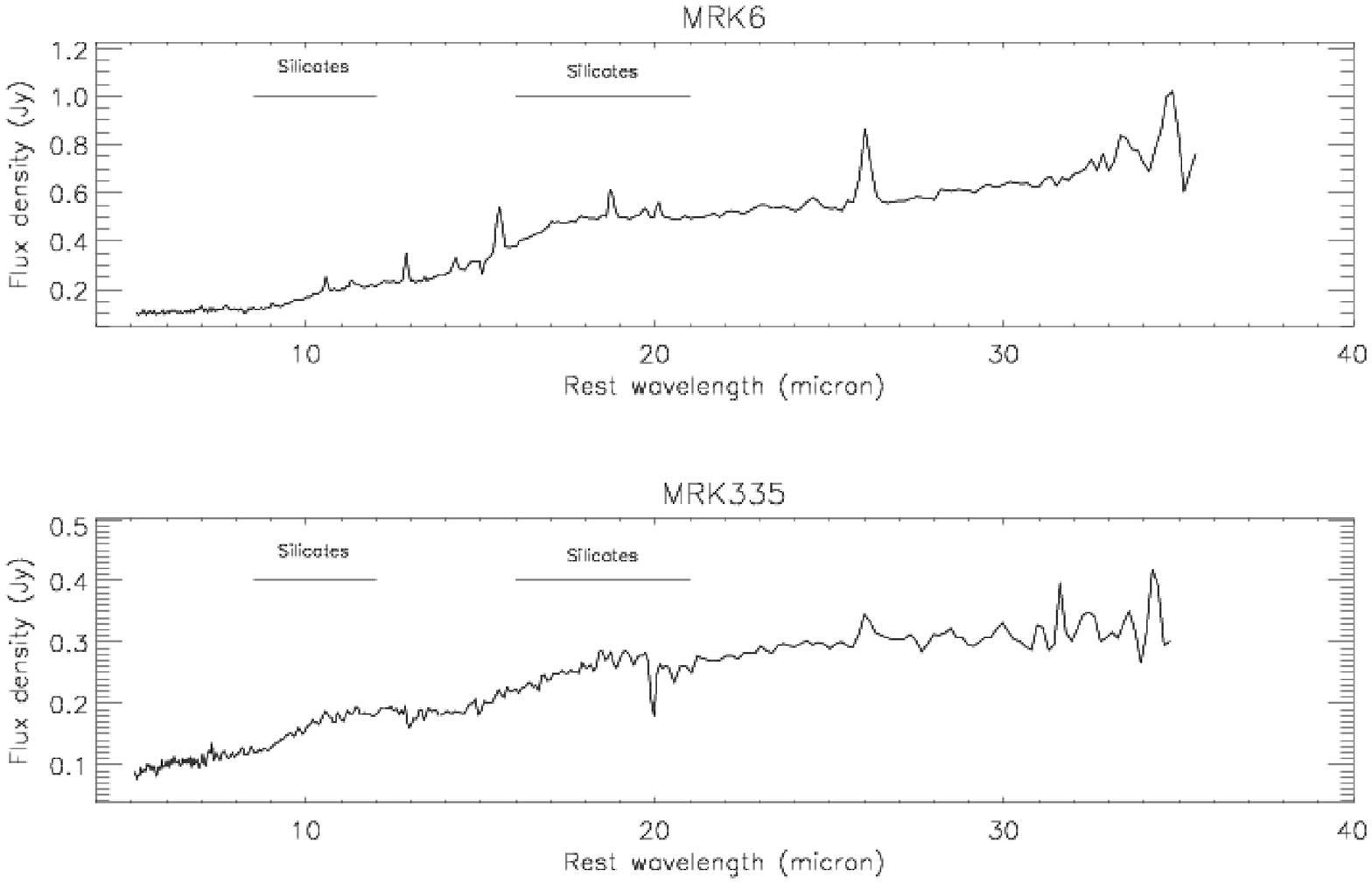}
\caption{{\it (b)} The spectra of the two objects, Mrk6 and Mrk335, that show
  clear silicate emission features at 10 and 18~\micron. The spectra have been
  converted to restframe.  \label{fig:silspec}} \epsscale{1.0}
\end{figure}

\begin{figure}
\epsscale{1.1}
\plottwo{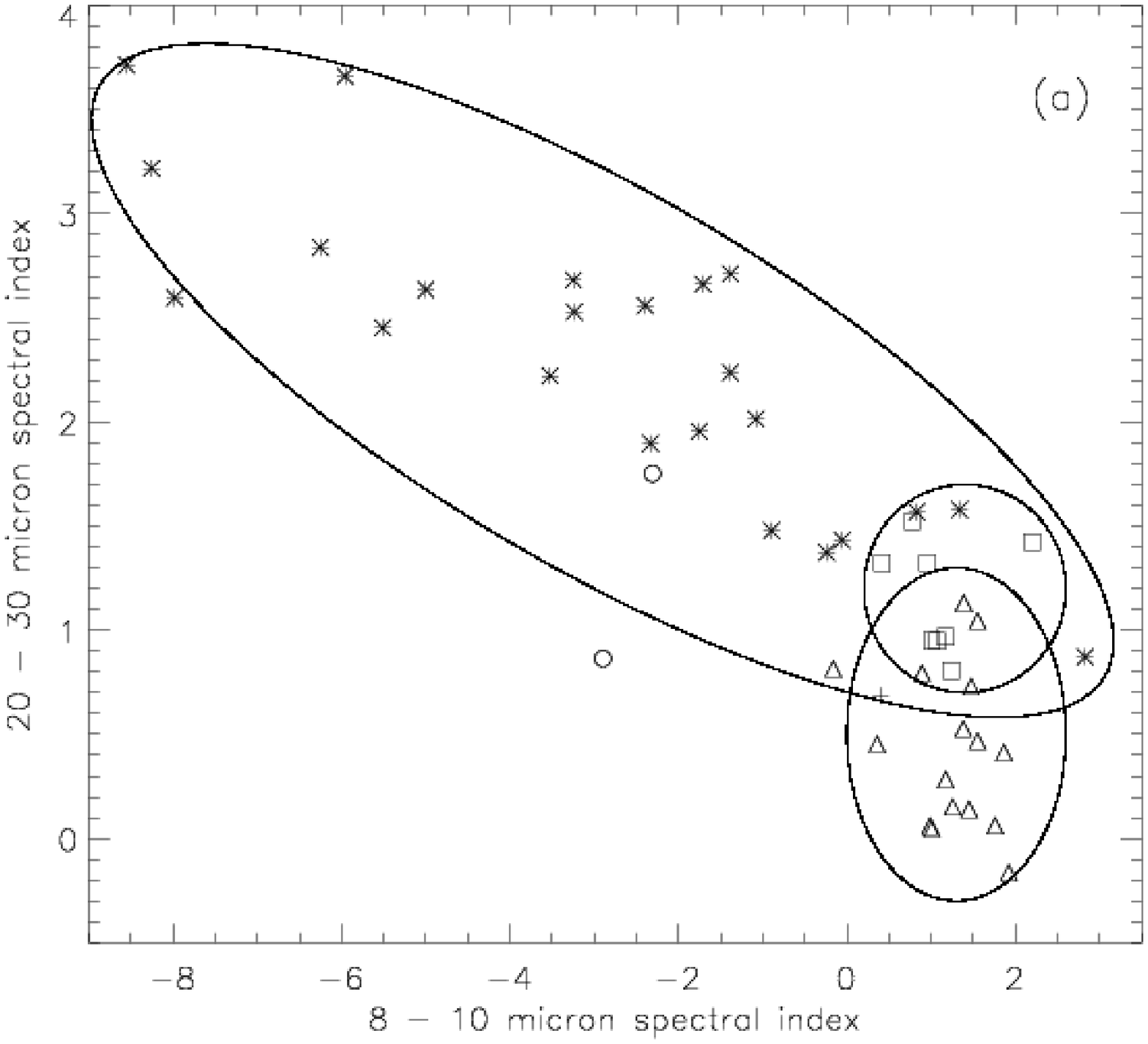}{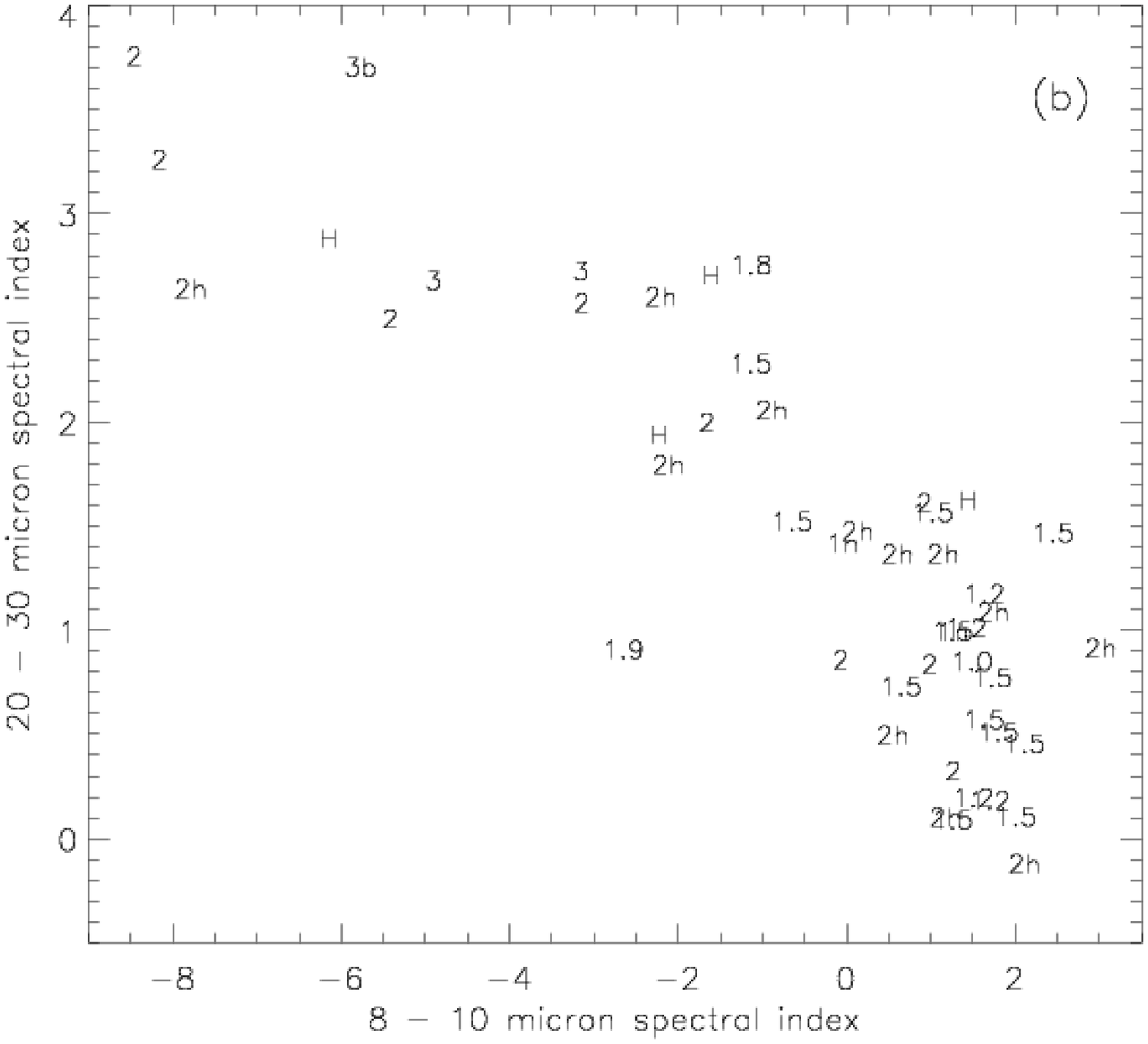}
\caption{IR color-color diagrams derived from the IRS spectra. {\it (a)} The
  symbols indicate the shape of the spectrum: red continuum with PAH features
  {\it (stars)}, broken power-law {\it (triangles)}, unbroken power-law {\it
  (squares)}, and silicate absorption {\it (circles)}. The ellipses show the
  areas occupied by the three largest groups; no ellipse is shown for the 2
  objects with silicate absorption. {\it (b)} The numbers indicate the Seyfert
  spectral type. Numbers 1.0 -- 1.9 and 1n represent Seyfert 1
  subtypes. Symbols 2h and 2 represent Seyfert 2's with and without HBLRs,
  respectively. LINERs, with and without broad lines, are shown by numbers 3
  and 3b, respectively, and starburst galaxies are indicated by the letter H.
  \label{fig:cols}} \epsscale{1.0}
\end{figure}

The spectral shapes may be quantitatively characterized by the IR colors
derived from the spectra. We calculate the spectral index between two
wavelengths using \[\alpha_{2-1} = \log(f_{\lambda_{1}}/f_{\lambda_{2}})
/\log(\lambda_{1}/\lambda_{2}).\] The spectral index between 20 and
30~\micron\ quantifies the slope of the spectrum (above the $\sim$20~\micron\
`break' for broken power-law spectra), while the spectral index between 8 and
10~\micron\ indicates the presence or absence of the 9.7~\micron\ silicate
feature and PAH features. The resulting color-color diagrams are given in
Figure \ref{fig:cols}.  One \sytwo\ (NGC3079) has extreme colors
$\alpha_{20-30\,\micron} = 4.7$ and $\alpha_{8-10\,\micron} = -12.5$, due to
its very red spectrum and strong PAH emission, so it is not shown in
Fig. \ref{fig:cols}.  The \syones\ (Seyfert types 1.0, 1.2 and 1.5) lie in the
lower-right part of the diagram, occupied by unbroken and broken power-law
spectra, specifically avoiding the region of the reddest PAH
emitters. \sytwos\ are found in both regions. Earlier type Seyferts (i.e. 1 --
1.5) have the bluest colors, while Seyfert 2's with hidden broad line regions
(HBLRs; type 2h) span the range of colors.  The fact that \syones\
consistently show the bluest infrared colors and the \sytwos\ are the reddest
sources indicates the IR spectra are changing systematically with Seyfert
type. We consider the implications further in \S~\ref{sec:dis}.

\clearpage

\subsection{Principal Component Analysis} \label{subsec:res_pca}

\begin{figure}
\epsscale{0.7}
\plotone{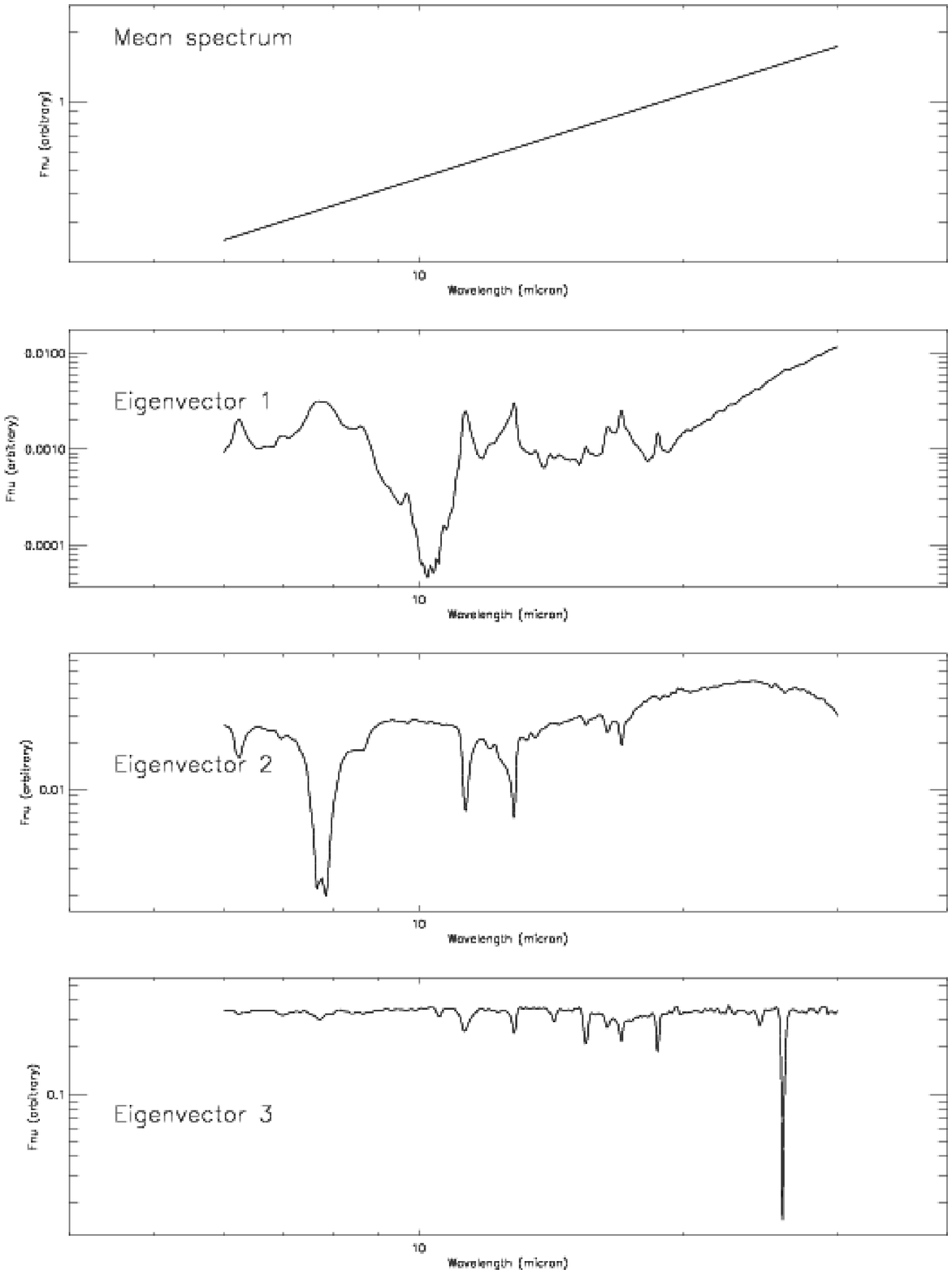}
\caption{The adopted `mean AGN' power-law spectrum {\it (top)} and first three
  eigenvectors of the sample of spectra, derived through principal component
  analysis.
  \label{fig:evectors}} \epsscale{1.0}
\end{figure}

We performed a principal component analysis (PCA) on the spectra to determine
the component spectral shapes (eigenvectors) producing the variety of spectra
seen in the sample (see, e.g., \citealt{fra99, sha04}). Because we are
presently more interested in the shape of the mid-infrared SED rather than
narrow emission features, we median-smoothed all of the spectra to improve the
continuous signal in the primary eigenvectors. We adopt a power law spectrum
with index $\alpha = -1.2$, (the average slope of the power-law spectra, where
$S_{\nu} \propto \nu^{\alpha}$) as the AGN `mean' spectrum, rather than using
the average spectrum derived from the data, to try to separate this component
of the spectra from other features, although using the actual mean spectrum
gives similar results.  The spectra were converted to the restframe and
normalized to unity flux density at 19.0~\micron\ before performing the
analysis.  The PCA computes $n$ eigenvectors, where $n$ is the number of
wavelength bins, but for the present data only the first few contribute
significantly to the variance. In fact, the first eigenvector is the most
dominant and contributes 91\% of the variance in the spectra.  The first 3
eigenvectors together account for 97.7\% of the variance, and the first 6
eigenvectors 99.0\%.  Figure \ref{fig:evectors} shows the power law spectrum
and the first three eigenvectors.

\begin{figure}
\epsscale{0.6}
\plotone{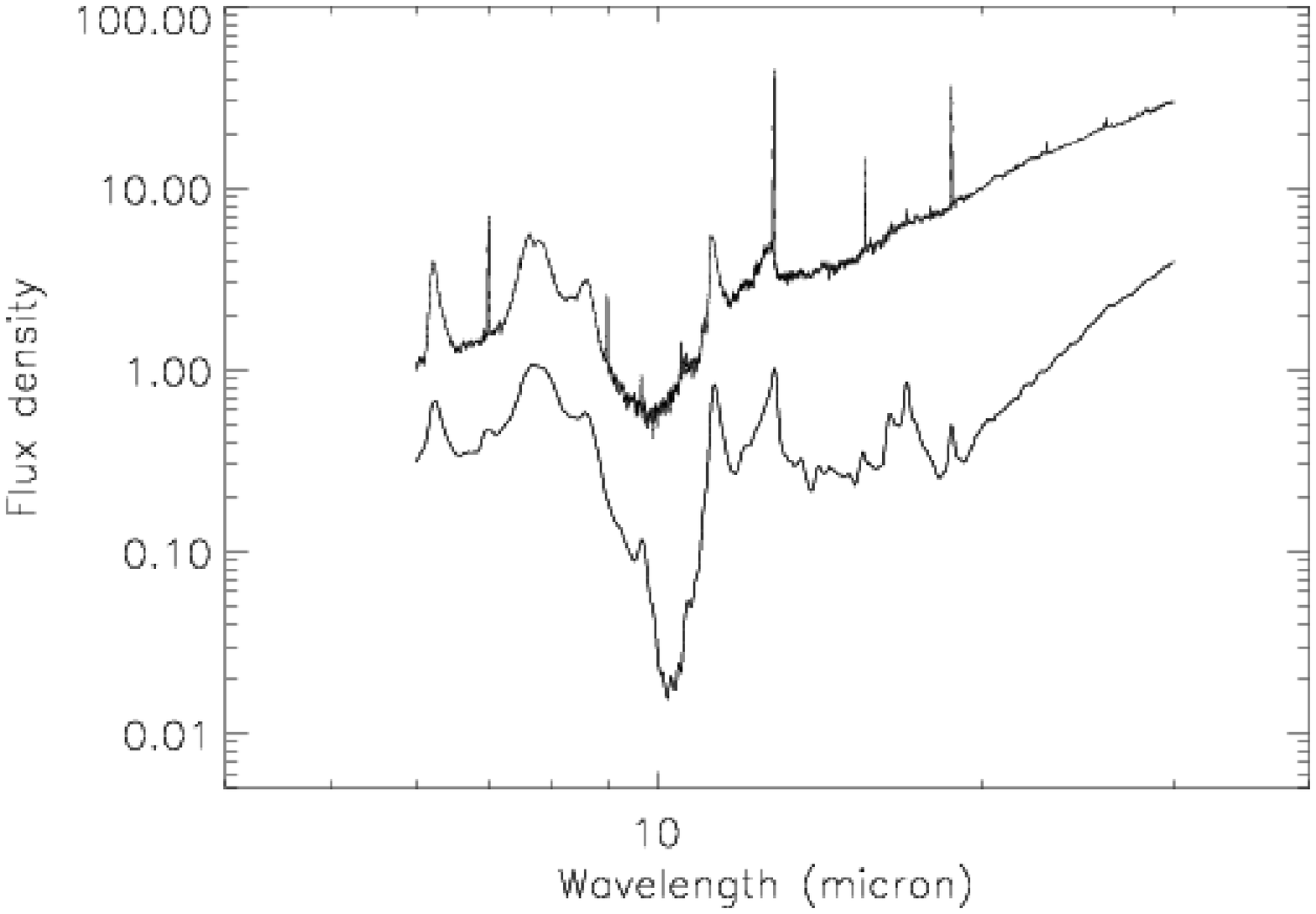}
\caption{The first eigenvector of the PCA which we interpret as dominated by a
  starburst component but also including contribution from other  features
  {\it (thin line)}, and, for comparison, the ISO spectrum of starburst galaxy
  M82 {\it (thick line)}. The ISO spectrum was obtained from
  http://isc.astro.cornell.edu/~sloan/library/swsatlas/atlas.html. The
  eigenvector clearly shows the PAH features and silicate absorption seen in
  starburst galaxies. The shape of the eigenvector at the longer wavelength
  differs slightly from a starburst shape. It is possible that this shape in
  the eigenvector produces the break in slope in the NGC~4151-like
  objects. \label{fig:sbcompare}} \epsscale{1.0}
\end{figure}

The first eigenvector shows broad emission features, a deep absorption feature
at $\sim$10~\micron, and red continuum above 20~\micron.  We interpret the
first eigenvector to be dominated by a starburst component, as it closely
resembles the spectra of starburst galaxies (Fig. \ref{fig:sbcompare}). The
emission features are attributable to PAH emission, while the absorption
feature is due to silicate dust absorption accentuated by the surrounding PAH
emission.  The change in continuum slope at $\sim$20~\micron\ in the PCA
eigenvector 1, which is not seen in starburst spectra, probably arises from
the broken power-law spectra in the sample, which have negative weights for
this eigenvector, and suggests that the first eigenvector does not represent a
`pure' starburst.  The second eigenvector is dominated by PAH features and so
accounts for differing relative strengths of the PAH emission with respect to
the silicate absorption in the spectra.  The third eigenvector appears to show
narrow line emission, in particular \oiv, indicating that the smoothing did
not completely remove these lines from the spectra.

\begin{figure}
\epsscale{0.4}
\plotone{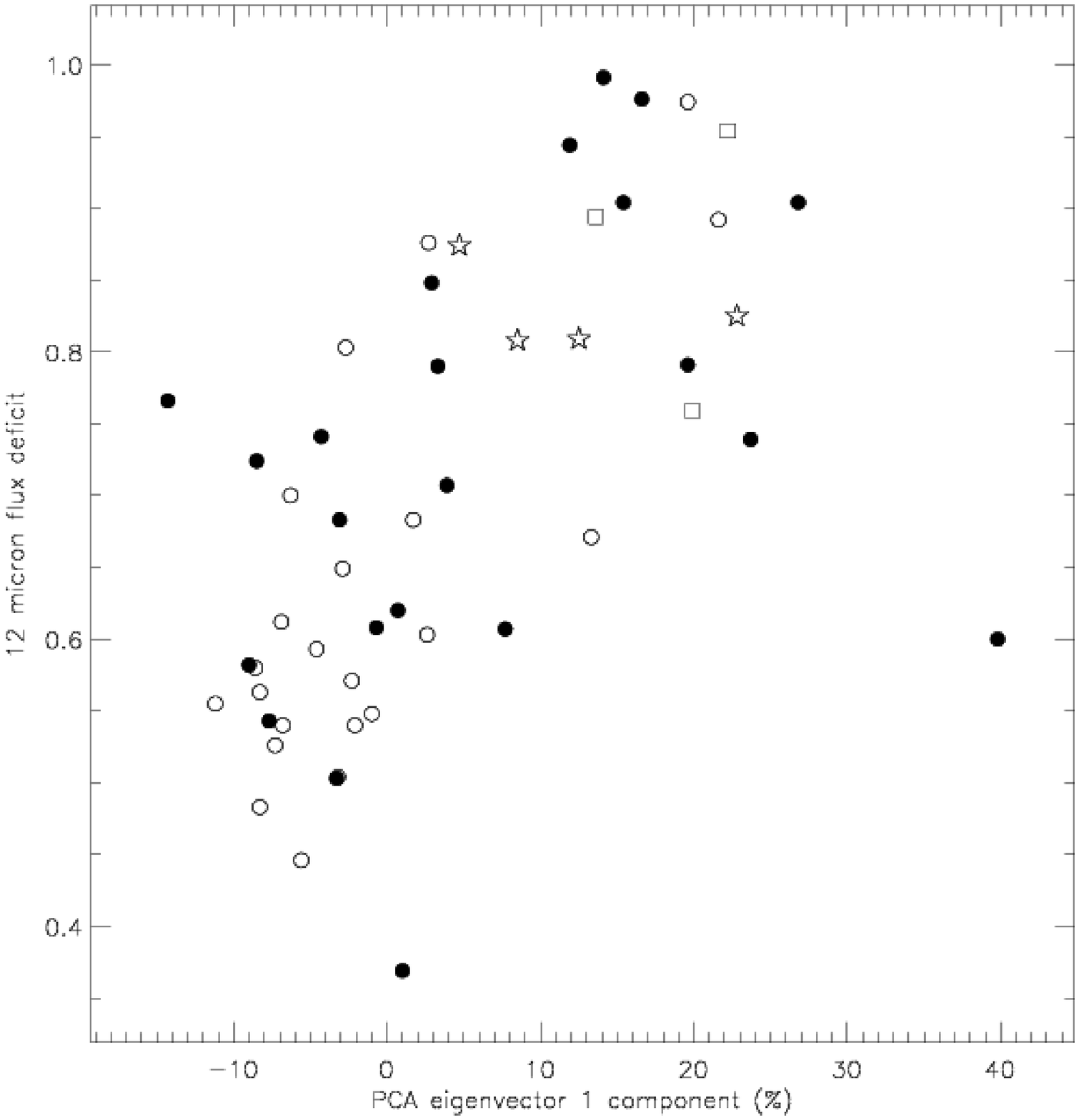}
\caption{The flux deficit at 12~\micron\ (see
  \S\protect\ref{sec:iras}) compared with the relative contribution of
  the first eigenvector derived from the PCA. The symbols indicate the
  Seyfert type: \syone\ {\it (open circles)}, \sytwo\ {\it (filled
  circles)}, LINER {\it (open square)}, and starburst galaxy {\it
  (open star)}.
  \label{fig:evonemiss}} 
\epsscale{1.0}
\end{figure}

\begin{figure}
\epsscale{0.5}
\plotone{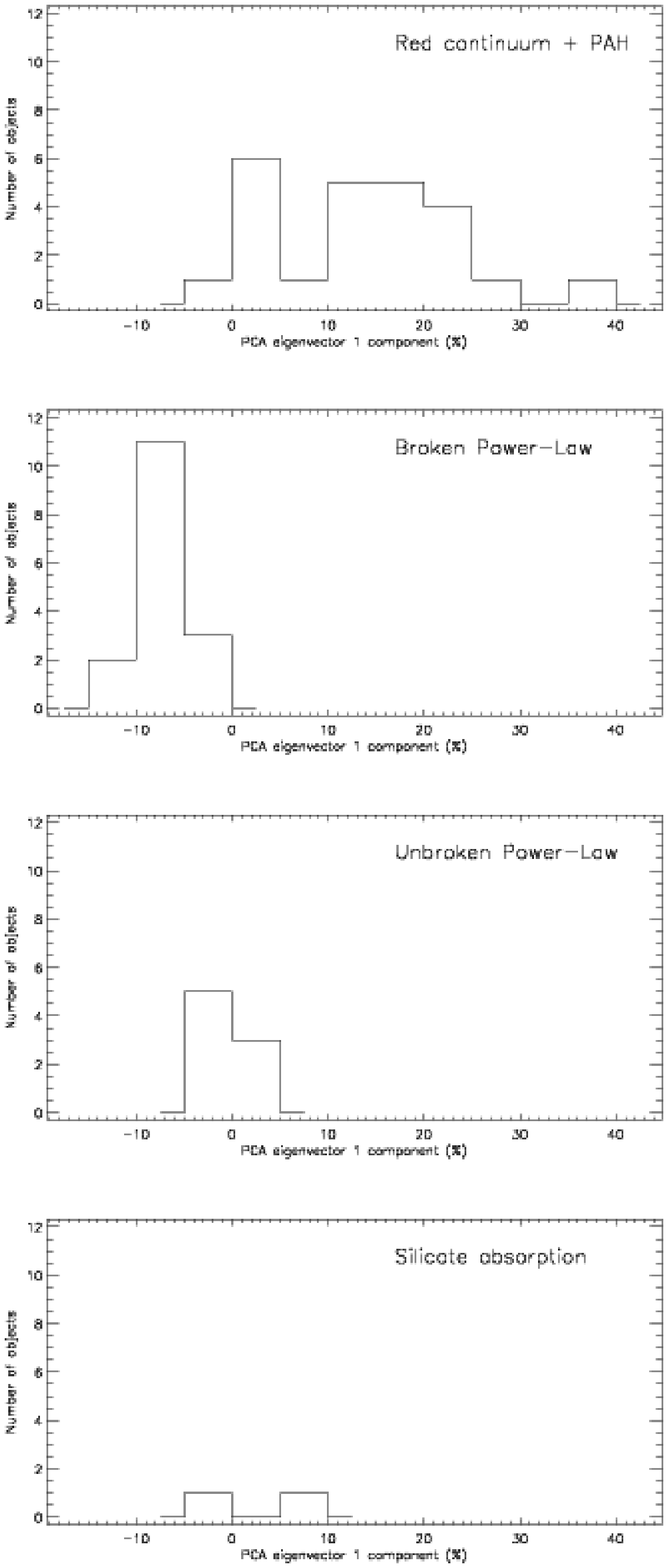}
\caption{Distribution of the relative contribution of eigenvector 1
  for the different shapes of the mid-infrared SED in the sample.
  \label{fig:evonehist}} \epsscale{1.0}
\end{figure}

The individual spectra, or component parts thereof, can be reproduced by
summing the `mean' spectrum and the eigenvectors multiplied by their
respective weights (e.g., \citealt{sha04}).  The absolute and relative
contributions of each eigenvector to each spectrum can be determined using the
associated weights. The relative contribution of the first eigenvector to each
spectrum appears to be related to the 12~\micron\ flux deficit (Figure
\ref{fig:evonemiss}), supporting the conclusion that it predominantly
indicates the starburst component of the SED.  This figure further suggests
that objects with more {\it extended} starburst activity also have more {\it
nuclear} starburst activity.  Figure \ref{fig:evonehist} shows that the
relative contribution of eigenvector 1 to each spectrum relates strongly to
the shape of the SED, as the different shapes have significantly different
contributions from this eigenvector (probabilities $<$0.03\% that the
distributions are from the same parent distribution, see Table
\ref{tab:statsb}).  Negative relative contributions indicate that this
component is subtracted from the `mean' spectrum to produce the observed
spectrum. Negative eigenvector 1 components in broken and unbroken power-law
spectra may incorporate 10~\micron\ silicate dust emission.  The spectra with
red continuum and PAH emission features show the largest eigenvector 1
components, consistent with these objects having the greatest starburst
contribution.

\begin{figure}
\epsscale{0.6}
\plotone{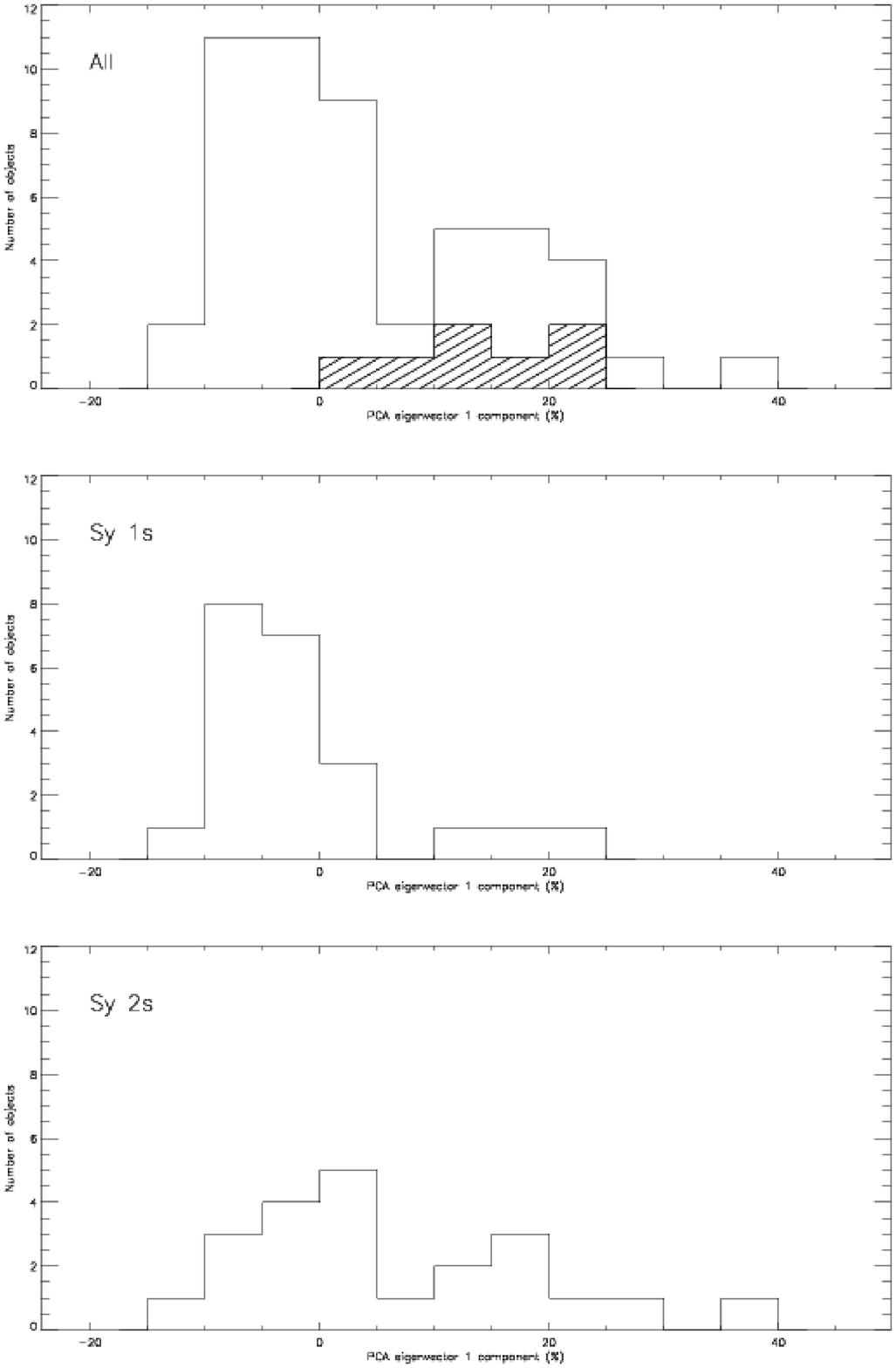}
\caption{Distribution of the relative contribution of eigenvector 1 for all
  observed objects {\it (top)}, \syones\ {\it (middle)} and \sytwos {\it
  (bottom)}. The hashed region in the top panel indicates the objects
  reclassified as non-Seyferts (LINERs and starburst galaxies).
  \label{fig:evonesy}} \epsscale{1.0}
\end{figure}

The relative contribution of the first eigenvector differs between Seyfert
types 1 and 2, with \syones\ having a negative median component and \sytwos\
having a positive median contribution.  This supports the result shown in
Figure \ref{fig:cols} that the spectra showing red continua and PAH emission
tend to be \sytwos\ while the \syones\ tend to have other shapes. However,
Figure \ref{fig:evonesy} also shows that the Seyfert types have overlapping
distributions of the first eigenvector contribution. The differences in the
first eigenvector contribution are less significant for Seyfert type than for
SED shape.  Consistent with this is the lack of separation of Seyfert types in
Figure \ref{fig:evonemiss}.  The statistical probabilities determined from
two-sample tests are listed in Table \ref{tab:statsb}.  We are currently
performing radiative transfer modeling of the dust emission in order to
determine the physical mechanisms producing the different IR shapes and
compare the Seyfert types in more detail (Buchanan et al., in prep).


\clearpage

\subsection{IR/Radio flux ratios} \label{subsec:res_rats}

\begin{figure}
\epsscale{0.40}
\plotone{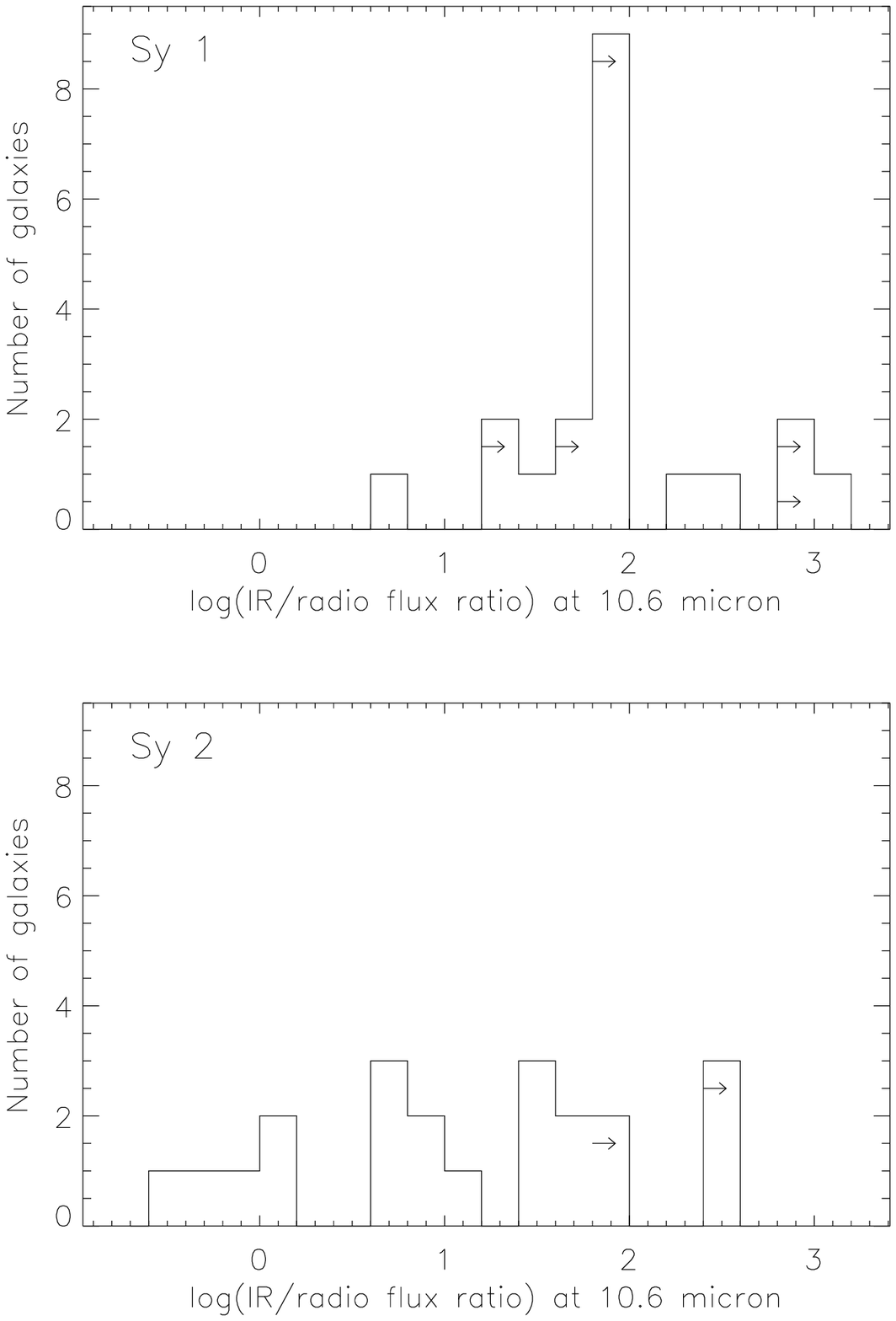}
\caption{Histograms of the 10.6~\micron\ to 8.4~GHz flux density ratio for the
  \protect\syones\ {\it (top)} and \protect\sytwos\ {\it (bottom)}. Lower
  limits (sources where the radio flux is an upper limit) are indicated by
  arrows.  \label{fig:ratiohist}} \epsscale{1.0}
\end{figure}

The unified scheme for active galaxies attributes the differences between
\syones\ and \sytwos\ to the presence of an optically thick dusty torus
surrounding the nucleus (see reviews by \citealt{ant93, urr95}). Models of
such a torus predict that, if the dust is optically thick at mid-IR
wavelengths, \syones\ (face-on) will be stronger mid-IR emitters than \sytwos\
(edge-on) \citep{pie92, gra94}.  \citet{hec95a} tested this prediction by
comparing the mid-IR emission of a heterogeneous sample of Seyfert
galaxies. He derived the ratios of 10.6~\micron\ to non-thermal radio flux
densities, to normalize the IR emission from each object with respect to its
intrinsic AGN brightness, and compared the IR/radio flux density ratios of the
type 1 and type 2 objects. As noted in \S\ref{sec:sample}, the optically thin
nuclear radio emission associated with the AGN is isotropic and considered to
be representative of the intrinsic AGN power.  \citet{hec95a} found that
\syones\ are indeed stronger 10.6~\micron\ emitters than \sytwos, relative to
their isotropic ``intrinsic'' AGN brightness, by a factor $\sim$4. This result
extended the work of \citet{mai95b}, who found that the \syones\ have brighter
absolute nuclear IR emission than \sytwos, showing that the difference is due
to anisotropy in the mid-IR emission.  We derive the nuclear IR to radio flux
density ratios of our homogeneous sample of Seyfert galaxies, using our IRS
spectra and the nuclear radio flux densities (Fig. \ref{fig:radhist}).  Figure
\ref{fig:ratiohist} shows the distributions of the ratio
$S_{\rm{10.6\,\micron}}/S_{\rm{8.4\,GHz}}$ for the \syones\ and \sytwos.  We
find the median of the \syones\ to be higher than that of the \sytwos, by a
factor $\sim$5 (see Table \ref{tab:statsb}).

\begin{figure}
\epsscale{1.0} 
\plottwo{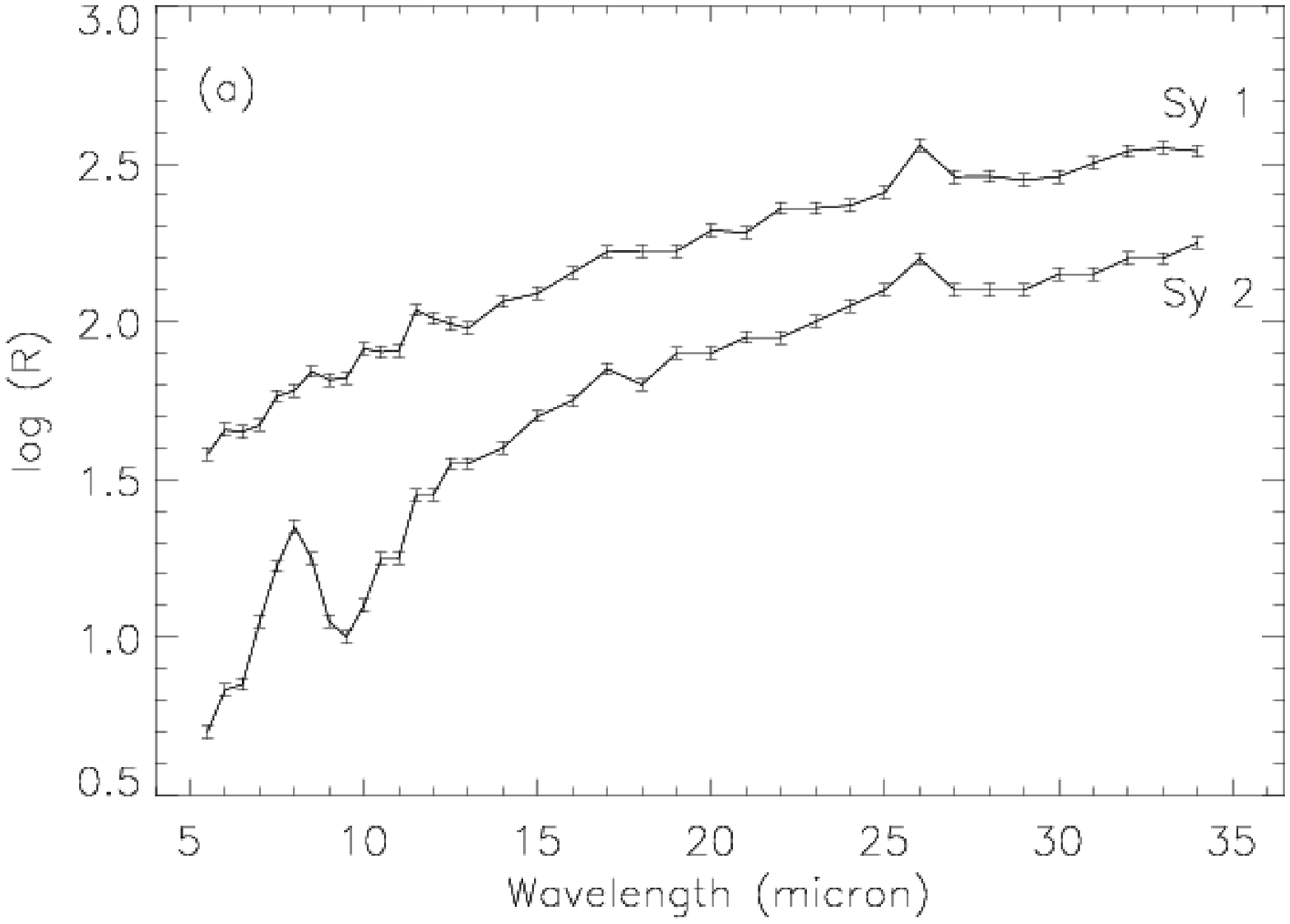}{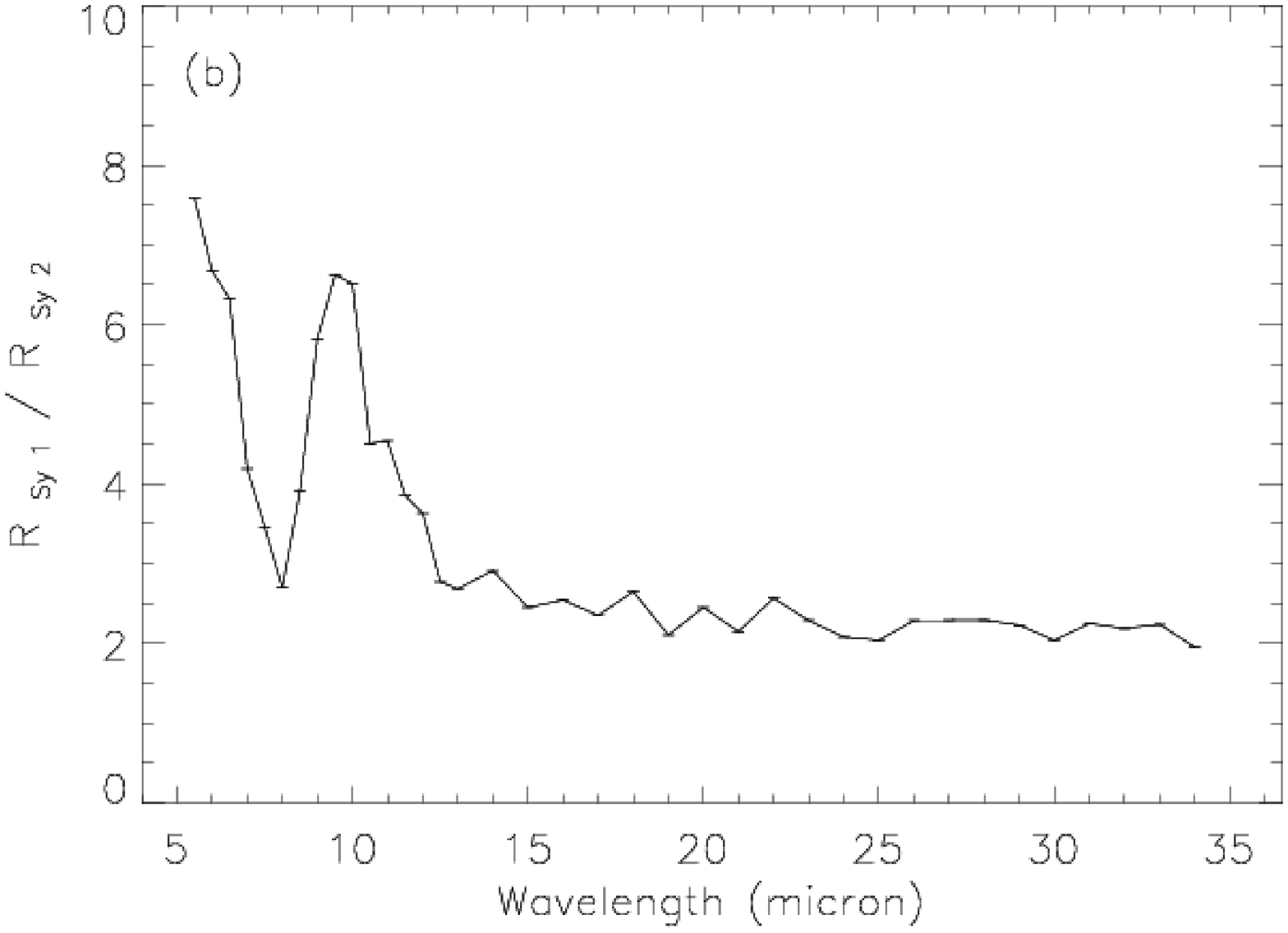}
\caption{{\it (a)} Logarithmic ratio log(R), where R = median IR/radio flux
  density ratio, of the \protect\syones\ and \protect\sytwos, as a function of
  the IR wavelength.  {\it (b)} The ratio of R for \protect\syones\ to that of
  \protect\sytwos\ at each IR wavelength. \label{fig:irradio}}
  \epsscale{1.0}
\end{figure}

We repeated this analysis at intervals of 0.5 -- 1.0~\micron\ along the IRS
spectral range, in order to investigate this difference between \syones\ and
\sytwos\ across the mid-IR. The median ratio of
$\log(S_{\rm{IR}}/S_{\rm{8.4\,GHz}})$ is consistently higher for \syones\ than
for \sytwos\ at all mid-IR wavelengths (Figure \ref{fig:irradio}), though the
statistical significance of the difference decreases at longer wavelengths.
The difference between \syones\ and \twos\ is particularly apparent around
10~\micron. This may be due to strong PAH emission and/or silicate absorption
in the \sytwos\ that is absent in the \syones. Figure \ref{fig:irradio} shows
the ratio of $S_{\rm{IR}}/S_{\rm{8.4\,GHz}}$ of \syones\ to that of
\sytwos. The sharp increase in this ratio towards shorter wavelengths might
indicate the presence of a hot dust component, peaking $\sim$2~\micron, in the
\syones\ that is absent in the \sytwos.  A higher ratio of IR/radio emission
in \syones\ is consistent with AGN unification, if the obscuring torus is
optically thick in the IR, however a clear separation of the starburst and AGN
components in the spectra, from detailed modeling, is necessary before further
interpretation of Figure \ref{fig:irradio} will be possible (see
\S\ref{subsec:dis_torus}). In particular, it is unclear if the 10~\micron\
peak arises in the torus (\sytwos\ being edge-on) or in the starburst
component (preferentially stronger in \sytwos).

\begin{figure}
\epsscale{0.6}
\plotone{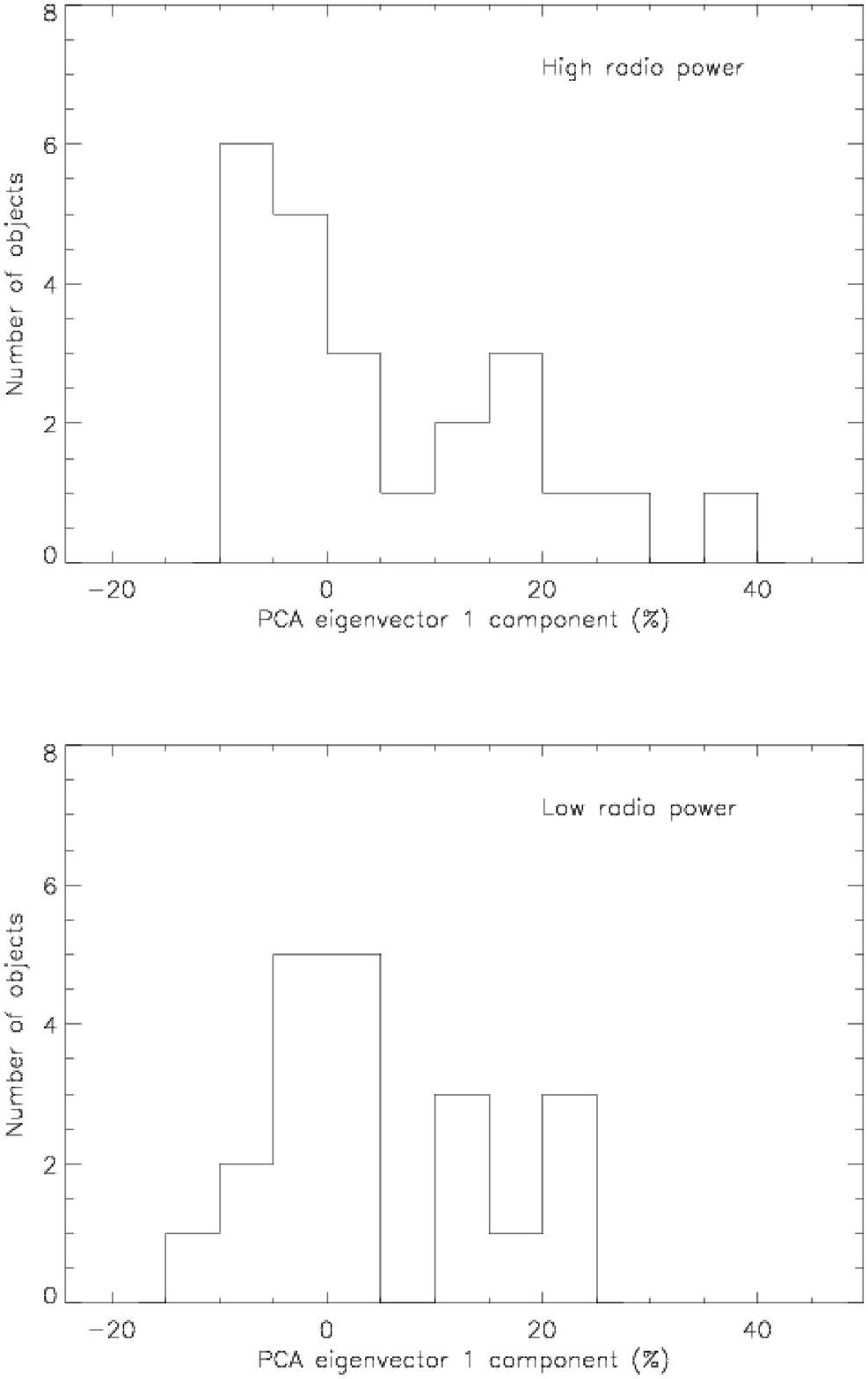}
\caption{Distribution of the relative contributions of eigenvector 1 for high
  {\it (top)} and low {\it (bottom)} radio power sources, separated by the
  median radio power of 2.2$\times 10^{21}$~\whz.  \label{fig:radpowerhist}}
  \epsscale{1.0}
\end{figure}

\clearpage

\section{Discussion} \label{sec:dis}

We now consider the implications of our results for our understanding of the
starburst/AGN connection, the AGN unified model, and selection effects in AGN
samples.

\subsection{The relative IR contributions of starburst and AGN} \label{subsec:dis_sbagn}

Spitzer spectra of the sample reveal distinct continuum shapes and dust
features. The similarity in the features between the red objects like MRK~938
(see Fig.\ \ref{fig:egspec}) and the IR spectra of starburst galaxies such as
M~82 (see Fig.\ \ref{fig:sbcompare}) strongly suggests these objects are
dominated by the starburst contribution to the dust heating at mid-infrared
wavelengths.  We find no difference in the redshifts of these objects compared
with the other IR spectral types, indicating that the relative strength of the
starburst component is not an aperture effect.  We do find evidence, however,
that the strength of the nuclear starburst, as measured by the contribution of
the first eigenvector in the PCA, is related to the extended star formation,
measured by the 12~\micron\ flux deficit (Figure \ref{fig:evonemiss}).  In the
starburst-dominated objects the dust is cool and so peaks longward of
30~\micron, and the PAH features are strong.  In contrast, objects with broken
and unbroken power-law spectra are probably dominated by the AGN contribution
to the spectrum.  Previous studies suggest that this AGN contribution is a
warm dust component (e.g., \citealt{rie78, mca88, bar87, pie93, pol00}).  It
is possible that the power-law spectra are produced by dust with a range of
temperatures \citep{pan75}, and the apparent break in the broken power-law
spectra is produced by a dominant dust component peaking at $\sim$20~\micron.
MFIR photometric SEDs of several Seyferts, based on {\it ISO} data, are
consistent with such a dust component \citep{rod96,per01}.  That the mid-IR
emission is dominated by dust is also supported by the recent identification
of silicate dust emission features in AGN mid-IR spectra
\citep{wee05,hao05c,sie05}.  In order to understand the spectral shapes, we
have begun comprehensive modeling using both AGN and starburst dust radiative
transfer models, the results of which will be reported elsewhere (Buchanan et
al., in prep.). The sharpness of the break in some IRS spectra near
20~\micron\ (e.g., NGC~4151; Fig.\ \ref{fig:egspec}) is difficult to reproduce
using simple dust models, however, even with several dust components.  The IR
emission is clearly produced by dust emission in the two objects with strong
silicate absorption at 10~\micron.

An alternative possibility is that the infrared emission may be non-thermal in
origin, such as synchrotron emission. The broad-band photometric SEDs of
nearby Seyfert galaxies provide evidence that the mid-IR emission of many
\syones\ is predominantly non-thermal \citep{ede87, war87, car87}, although a
non-thermal IR component is not always necessary to model AGN SEDs
\citep{bar90}. The detection of rapid variability at 10~\micron\ in some
nearby quasars indicates a non-thermal component of the mid-IR emission
\citep{neu99}, and further supports that the \syones\ may also have a
non-thermal IR contribution.  Comparison of the IR and nuclear 8.4~GHz radio
flux densities suggests that the infrared emission is not a continuation of
the radio synchrotron spectrum, as the IR fluxes are at least an order of
magnitude too high. If the synchrotron were self-absorbed and the spectral
turnover occurred at 20~\micron, consistent with the observed break in the
power-law spectra, the component would have to be $\lesssim 10^{11}$~m in
size, on the same scale as the accretion disk \citep{urr95}.  The typical
spectral slope at frequencies below (i.e., wavelengths above 20~\micron) the
turnover would have to be 0.8 to match the radio flux densities. This slope,
and the typical slopes at frequencies above the turnover (wavelengths below
20~\micron) of $\sim$-1.3, are not inconsistent with self-absorbed synchrotron
emission, so we cannot rule this out as the origin of the power-law
spectra. However, the energy density of such a synchrotron component would be
very large ($\sim$9$\times$10$^{8}$~erg\,cm$^{-3}$), so extreme conditions
would be required to confine it. The total energy in the plasma would also be
large, ($\sim$10$^{40}$~ergs), suggesting that such a component is unlikely.

A third alternative for the broken and unbroken power-law spectra is free-free
bremsstrahlung, however this would produce a much flatter slope ($S_{\nu}
\propto \nu^{-0.4}$) \citep{bar93} than is observed in the spectra.

A final possibility is that the IR emission is a continuation of the accretion
disk spectrum, with a slope of, for example, 1/3 (Eq. 5.46, p.92;
\citealt{fra92}), though more realistic accretion models will modify this
(e.g., \citealt{lit89}). This will be investigated in the future using X-ray
-- optical data in conjunction with the IR data.

The results of our principal component analysis show that the first
eigenvector produces more than 90\% of the variance in the observed spectral
shapes and is dominated by a starburst spectral shape. This suggests that the
relative contribution of star-formation to the IR spectrum is an important
factor determining the observed spectral shape. However, we emphasize that,
while the starburst contribution dominates the observed differences in the IR
spectral shapes, the starburst component does not dominate the total emission
of each object. Most spectra are comprised of a power law plus a $<$20\%
contribution of eigenvector 1 (Figure \ref{fig:evonesy}). We found a
significant difference between the contribution of the first eigenvector to
the spectra of the \syones\ and \sytwos.  This is consistent with the (on
average) higher flux deficits at 12 and 25~\micron\ in \sytwos\ compared with
\syones, suggesting that extended flux (i.e., associated with circumnuclear
star formation) contributes more to the total mid-IR flux in \sytwos\ than
\syones.  Our results confirm the finding of \citet{mai95b}, based on
ground-based photometry, that \sytwos\ show more extended star formation than
\syones.  The sample of objects studied by \citet{mai95b} were optically
selected, and therefore our result is not due to an infrared selection bias.
We have concluded that the extended emission is associated with star
formation, however it has also been shown that extended IR emission, on scales
of tens of parsecs, can also be associated with the AGN (e.g.,
\citealt{mas06,boc98}).  Our data demonstrates a clear relationship between
the Seyfert type and the IR SED shape (i.e., \syones\ tend to have power-law
or broken power-law SEDs and \sytwos\ have greater starburst contribution to
their SEDs), but there is not a 1-to-1 correspondence between the optical
Seyfert type and the IR spectral shape.  It is clear that star formation and
complex dust obscuration play key roles in producing the observed variety of
spectral shapes in the IRS spectra.  Homogeneous optical spectra, to ensure
uniform Seyfert classifications, and simultaneous modeling of the optical and
IR SEDs are necessary to elucidate these relations.

\subsection{Properties of the obscuring medium (comparison with torus models)}
\label{subsec:dis_torus} 

Comparing the optical spectral types of the IR spectral groups
(\S\ref{sec:res_spe}), it is not immediately apparent whether or not the IR
spectra support orientation-dependent obscuration.  Torus models predict
silicate emission in type 1 (face-on) objects and silicate absorption features
in type 2 (edge-on) objects (\citealt{row89,pie92,sie04}, though see also
\citealt{nen02}).  We would expect edge-on objects to show the most dust
absorption.  The two objects that show strong silicate absorption have optical
spectral types Sy~1.9 and HBLR \sytwo, which is not inconsistent with the
unified scheme. The majority of broken power-law spectra that may show
silicate emission features are Sy~1.2 and Sy~1.5, although the group includes
several \sytwos\ as well.  The relative contribution of a starburst to the IR
spectrum appears to be greater in \sytwos, possibly due to a selection effect
(see \S\ref{subsec:dis_sel}), so that it is difficult to discern any
orientation-dependent difference in the AGN component of the spectra.

The ratio of IR to radio flux densities is consistently higher for the type 1
objects in our sample than the type 2 objects, at all observed wavelengths.
Interpreting this in the context of the unified scheme, this may indicate that
the obscuring material is optically thick up to at least 30~\micron.  However,
optically thick smooth-density torus models predict much larger differences
between the face-on and edge-on Seyferts than the observed factor of $\sim$2,
seen in our data at wavelengths longer than $\sim$15~\micron\ (Figure
\ref{fig:irradio}). For example, \citet{pie92} show that, for an optically
thick torus, the face-on objects are expected to be at least a factor of 5 and
up to several orders of magnitude brighter than the edge-on objects.
\citet{lut04} first noted that torus models cannot explain the observed lack
of anisotropy in the IR emission of Seyferts, when they compared the
6~\micron\ emission of \syones\ and \twos, normalized to the intrinsic hard
X-ray emission.  Our findings, while not identical to those of \citet{lut04}
at 6~\micron\ (discussed below), confirm that observational data do not match
the predictions of optically thick smooth-density torus models.

Our results are consistent with the torus becoming optically {\it thin} in the
mid-IR, at around 15~\micron.  However, a mid-IR optical depth $<$1 would
imply column density $< 10^{23}$~\cms\ (e.g., \citealt{lut04}), which is
inconsistent with X-ray observations of many Seyfert 2 galaxies (e.g.,
\citealt{bas99}).  The observed high X-ray column densities of \sytwos\ could
be reconciled with optically thin IR emission, if the X-ray absorption is
dominated by gas in which the dust has been destroyed, e.g., close to the
nucleus within the dust sublimation radius. There is some evidence that this
may be possible: variability in the X-ray column densities of two Seyfert
\twos\ on timescales of less than a day indicates that absorbing columns of up
to $5 \times$ 10$^{23}$~\cms\ exist on scales of less than a few hundred
Schwarzchild radii \citep{ris05,elv04}. Thus our IR/radio flux density ratios
could indicate that the torus is optically thin above 15~\micron.  As noted in
\S\ref{subsec:res_rats} the difference in the ratio of IR/radio flux densities
between \syones\ and \sytwos\ is particularly apparent around 10~\micron; the
10~\micron\ flux density from an optically thin torus is predicted to show
more anisotropy than the surrounding continuum \citep{pie92}, suggesting that
our results are also consistent with an optically thin torus at shorter mid-IR
wavelengths than 15~\micron. However, it is not clear whether the 6.2~\micron\
PAH feature associated with star formation is contributing to the apparent
peak at 10~\micron\ in Figure \ref{fig:irradio}{\it (b)}.  Certainly, the
steep rise in this figure towards the shortest wavelengths (from 7 down to
5~\micron) suggests that the obscuring material is optically thick at these
wavelengths.

Clumpy torus models can accommodate much less anisotropy of the mid-IR
emission than models with smooth radial density distributions (e.g.,
\citealt{nen02}).  The torus emitting near-isotropically in the mid-IR, but
with anisotropic X-ray absorption, can be explained by clumpy torus models, if
the radial distribution of optically thick dust clumps is steep (Nenkova et
al.\ 2006, in preparation).  The similar ratios of IR/radio flux density of
\syones\ and \sytwos\ observed in our sample therefore may be explained by an
optically thick clumpy torus.

Our IR/radio ratios at 6~\micron\ appear to be in disagreement with those of
\citet{lut04}, who found no significant difference between \syones\ and
\sytwos\ in their ratio of X-ray/IR flux densities. These authors used the
intrinsic (i.e., absorption-corrected) X-ray flux densities as a measure of
the AGN luminosity, and determined the ratio of the X-ray flux density to the
6~\micron\ AGN continuum flux density, derived from ISO spectra. \citet{lut04}
find the average X-ray/IR flux ratio of \syones\ in their sample to be a
factor of $\sim$2 {\it less} than the average ratio for type 2 Seyferts, which
is contrary to the expectations of torus models in the unified scheme.  Our
results show a factor of $\sim$7 difference at this wavelength
(Fig. \ref{fig:irradio}), in the opposite direction.  There are several
factors which are likely to contribute to the disagreement between our results
and those of \citet{lut04}.  Firstly, the relative AGN and starburst
contributions to the IR flux densities are likely to be different in the two
studies. The ISO data of \citet{lut04} were obtained using a 24\arcsec\
aperture, which will include a large contribution of extended emission; the
authors account for this using spectral decomposition to determine the AGN
continuum.  In contrast our data are relatively small aperture and so better
isolate the nuclear emission, but we have not attempted to remove the
starburst contribution to the IR emission, which is clearly different for the
type \ones\ and \twos\ in our sample.  Secondly, the sample studied by
\citet{lut04} was heterogeneous and includes, for example, \syones\ and
\sytwos\ with different distance and intrinsic luminosity distributions. The
redshift and intrinsic (radio) luminosity distributions of the type \ones\ and
\twos\ in our homogeneous sample are similar (see \S\ref{sec:sample}) and so
will not have biased the comparison of the mid-IR emission.  Finally, the
studies are using measurements at different wavelengths to represent the
intrinsic AGN luminosities.  Large corrections (up to an order of magnitude)
are required to determine the intrinsic X-ray flux densities from the observed
flux densities, and therefore there may be considerable uncertainty in the
obtained intrinsic AGN luminosities, although we note that \citet{lut04} find
no evidence for a systematic offset in the absorption-corrected values which
would bias their result.  Given the various systematic biases in each of the
studies, it is perhaps not surprising that the exact ratios of \syone/\sytwo/
IR emission measured are not the same.  However, despite this disagreement,
our results at the longer wavelengths lead us to the same conclusion as
\citet{lut04}: the data are not consistent with optically thick,
smooth-density torus models (e.g, \citealt{efs95}), which predict strongly
anisotropic mid-IR emission.

\subsection{Selection effects in the IR} \label{subsec:dis_sel}

In the context of the unified scheme, the IR/radio flux density ratios of the
\syones\ and \sytwos\ in our sample suggest that the torus may be optically
thick at wavelengths shorter than $\sim$15~\micron\ (see Figure
\ref{fig:irradio} and \S\ref{subsec:dis_torus}).  The Seyfert sample was
selected using IRAS fluxes at 12~\micron, therefore we might expect there to
be orientation-dependent selection effects biasing our sample, that is, that
dust obscuration will bias the sample towards type \ones. A second possible
selection effect, if \sytwos\ have more star formation than \syones, is
discussed below.  To fall above the flux limit of the sample, the \sytwos\
must contain intrinsically brighter AGNs than the \syones, because the
\sytwos\ suffer more MIR dust obscuration.  However, we find no significant
difference between the AGN luminosities of the \syones\ and \sytwos\ in the
sample (as measured by the optically thin radio emission, see
\S\ref{sec:sample}).  Further, if the \sytwos\ have more luminous AGN emission
than the \syones\ due to a selection effect, we may expect that the
contribution of the first eigenvector would be different for high and low
radio power sources. Again, we find no significant difference in the
distributions of relative eigenvector 1 components for high ($>10^{24}$~\whz)
and low power radio sources in the sample (Figure \ref{fig:radpowerhist}).  It
is possible that small number statistics are preventing us from detecting a
significant difference in the radio powers of the \syones\ and \twos\ (see
Table \ref{tab:statsa}).  The lower IR/radio flux density ratios observed in
\sytwos\ compared with \syones\ are consistent with more luminous radio
sources in \syones.

A second selection effect could be operating in the opposite direction to that
associated with orientation-dependent obscuration of the nucleus. \sytwos\
with intrinsically faint nuclei could make it into the sample if they have an
additional starburst contribution to the 12~\micron\ flux density. This would
result in the sample being biased towards \sytwos\ with a greater relative and
absolute starburst contribution than the \syones\ in the sample.  It is clear
from our color-color diagram (Fig.\ \ref{fig:cols}) and principal component
analysis that the observed \sytwos\ show a stronger starburst contribution to
their infrared spectra than the \syones, consistent with such a bias.  A
near-infrared study has shown that the \sytwos\ in the 12~\micron\ sample are
not biased towards those with powerful nuclear starbursts compared with
optically-selected \sytwos\ \citep{ima04}, but the 12~\micron-selected
\sytwos\ may be biased towards more luminous \emph{extended} starbursts due to
the large-aperture IRAS observations used to select them.

Thus we cannot exclude the possibility that the two selection effects
described above are canceling each other out to give the same distribution of
infrared luminosities for \syones\ and \twos.  However, it is puzzling that,
if the sample is biased towards \sytwos\ with intrinsically brighter AGN
and/or those with greater starburst contribution to their IR emission, that we
see no differences in the AGN (radio) luminosities.  The similar distribution
of intrinsic AGN luminosities (radio powers) may imply that the effects of
obscuration and starburst contribution are small compared with the AGN
luminosities.  Alternatively, the statistical results may become clearer when
more of our data are available. Further leverage on selection effects and
biases can be obtained using IRAC imaging data, which we will address in
future work.

The selection of Seyfert samples based on the isotropic luminosity of the AGN
and which are unbiased with respect to orientation is essential to study the
physics of active galaxies and the AGN/starburst connection.  An unbiased
sample is difficult to obtain, however.  The optical and 0.1 -- 10~keV X-ray
continuum fluxes suffer uncertain orientation dependent obscuration so favor
\syones\ and discriminate against \sytwos\ (e.g., \citealt{gre87, hec95b}).
Optical line emission, such as \oiii, originates in the narrow line region so
should be isotropic and representative of the intrinsic AGN luminosity.
However it has been shown recently \citep{hao05b} that, while the numbers of
\ones\ and \twos\ are comparable at low [\ion{O}{3}] luminosity, Seyfert
\ones\ outnumber the \twos\ by a factor of about 3 at high [\ion{O}{3}]
luminosity, suggesting that [\ion{O}{3}] luminosity may not be an unbiased
criterion. It has further been argued, for powerful AGN, that some
[\ion{O}{3}] emission may come from within the torus, and therefore be
orientation-dependent \citep{dis97}.  \citet{sm89} argued that selection
effects are minimal in the 7 -- 12~$\micron$ region.  The extended 12~\micron\
sample of Seyfert galaxies was thus selected to minimize wavelength-dependent
selection effects.  Our results have shown, however, that the nuclear
obscuring material may be optically thick at MIR wavelengths up to 15~\micron\
and so selection effects are still important in the mid-IR.  The torus
presumably becomes optically thin at far-IR wavelengths, but at these
wavelengths it becomes difficult to separate nuclear emission from
circumnuclear star formation, which emits strongly at FIR wavelengths due to
the presence of cold dust.  Similarly, at low radio frequencies and spatial
resolutions, such as the FIRST survey at 1.4~GHz \citep{whi97}, the relatively
strong contribution of star formation to the nuclear spectrum may bias the
selection.  To obtain a sample of Seyfert galaxies unbiased by possible
orientation effects, it is therefore necessary to select galaxies based on
optically thin, compact radio emission associated with the active nucleus
(e.g., \citealt{the01}).  A large area (10 square degrees), sensitive (0.2
mJy/beam) survey at high spatial resolution (0.2 \arcsec) at high frequency
(8~GHz) would seem to be a useful way to select nearby \syones\ and \sytwos\
unbiased by orientation effects.

\section{Conclusions} \label{sec:con}

We present the results of Spitzer IRS nuclear spectra for the first 51 of a
complete sample of 12~\micron\ selected Seyfert galaxies.  We find the
following main conclusions: \\
$\bullet$ The spectra clearly divide into groups based on their continuum
shapes and spectral features. The largest group (47\% of the sample of 51)
shows very red continuum suggestive of cool dust and strong emission features
attributed to PAHs. Sixteen objects (31\%) have a power-law continuum with
spectral indices $\alpha_{5-20\,\micron} = $-2.3 -- -0.9 that flatten to
$\alpha_{20-35\,\micron} = $-1.1 -- 0.0 at $\sim$20~\micron.  Clear silicate
emission features at 10 and 18~\micron\ are found in two of these objects
(Mrk~6 and Mrk~335).  A further 16\% of the sample show power-law continua
with unchanging slopes of $\alpha_{5-35\,\micron} = $-1.7 -- -1.1. Two objects
are dominated by a broad silicate absorption feature. One object in the sample
shows an unusual spectrum dominated by emission features, that is unlike any
of the other spectra. Some features are clearly related to the starburst
contribution to the IR spectrum, while the mechanisms producing the power-law
continuum attributed to the AGN component are not yet clear. The precise
correspondence between the IR types and the optical spectral types is unclear,
firstly because the exact contributions of the AGN and starburst to the IR
emission are not clearly separated by colors or principal component analysis,
and secondly because the Seyfert types are heterogeneous and often ambiguous.
\\
$\bullet$ Principal component analysis suggests that the relative contribution
of starburst-heated dust emission to the SED is the dominant cause of variance
in the observed spectra, although all the spectra are dominated by a power-law
component. The close detailed agreement of eigenvector 1 with the spectrum of
a starburst at $\lambda < $20~\micron\ makes it almost certain that
eigenvector 1 is tracking the starburst component in these AGNs. Moreover,
that eigenvector 1 is linked to 12~\micron\ IRAS/Spitzer flux deficit, shows
that this star formation is extended and so not a purely nuclear
phenomenon. \\
$\bullet$ We find that the \sytwos\ typically show stronger starburst
contributions in their IR spectra than the \syones, confirming previous
results found using photometric data. This result is in disagreement with the
predictions of unified schemes. The stronger starburst contribution in
\sytwos\ compared with \syones\ may be a selection effect in our sample,
though we find no evidence supporting the presence of a bias in our sample. \\
$\bullet$ We confirm the previous finding that \syones\ show higher ratios of
IR/radio emission than \sytwos\ at $\sim$10~\micron, by a factor $>$6, however
we find that the difference between type \ones\ and \twos\ decreases to a
factor of $\sim$2 beyond $\sim$15~\micron.  In the context of unified schemes,
this presents a challenge for dusty torus models, which generally predict much
larger anisotropy. The observed factor of $\sim$2 difference between the type
\ones\ and \twos\ in their IR/radio ratios above 15~\micron\ requires the
standard smooth-density torus models to be optically thin at these
wavelengths.  However, the resulting low torus opacity requires that the high
observed columns detected in X-ray absorption be produced in gas with very low
dust to gas ratio (perhaps within the dust sublimation region). On the other
hand, our observations may be consistent with clumpy torus models (e.g.,
\citealt{nen02}) containing a steep radial distribution of optically thick
dense clumps. \\
$\bullet$ The selection of our sample at 12~\micron, where the torus may be
optically thick, implies there may be orientation-dependent biases in the
sample, however we see no evidence for these in our results. \\
Detailed modeling of the continuum emission is underway to separate in detail
the starburst and AGN contributions to the IR spectrum in order to place
constraints on the opacity and geometry of the nuclear obscuring material, and
to compare the relative starburst contributions of Seyfert types 1 and 2.

\acknowledgments

We thank the referee for a careful reading of the manuscript, which
resulted in significant improvements to the paper. This work is based
on observations made with the Spitzer Space Telescope, which is
operated by the Jet Propulsion Laboratory, California Institute of
Technology under a contract with NASA. Support for this work was
provided by NASA through an award issued by JPL/Caltech.  The IRS was
a collaborative venture between Cornell University and Ball Aerospace
Corporation funded by NASA through the Jet Propulsion Laboratory and
Ames Research Center.  SMART was developed by the IRS Team at Cornell
University and is available through the Spitzer Science Center at
Caltech. This research has made use of NASA's Astrophysics Data
System.

{\it Facilities:} \facility{Spitzer}


\small
\begin{thebibliography}{}
\parskip -1mm
\bibitem[Antonucci(1993)]{ant93} Antonucci, R.\ 1993, \araa, 31, 473
\bibitem[Barvainis(1987)]{bar87} Barvainis, R.\ 1987, \apj, 320, 537
\bibitem[Barvainis(1990)]{bar90} Barvainis, R.\ 1990, \apj, 353, 419
\bibitem[Barvainis(1993)]{bar93} Barvainis, R.\ 1993, \apj, 412, 513
\bibitem[Bassani et al.(1999)]{bas99} Bassani, L., Dadina, M.,
Maiolino, R., Salvati, M., Risaliti, G., della Ceca, R., Matt, G., \&
Zamorani, G.\ 1999, \apjs, 121, 473
\bibitem[Bock et al.(1998)]{boc98} Bock, J.~J., Marsh, K.~A., Ressler, M.~E.,
\& Werner, M.~W.\ 1998, \apjl, 504, L5
\bibitem[Carleton et al.(1987)]{car87} Carleton, N.~P., Elvis, M., Fabbiano,
G., Willner, S.~P., Lawrence, A., \& Ward, M.\ 1987, \apj, 318, 595
\bibitem[di Serego Alighieri et al.(1997)]{dis97} di Serego Alighieri, S.,
Cimatti, A., Fosbury, R.~A.~E., \& Hes, R.\ 1997, \aap, 328, 510
\bibitem[Edelson(1987)]{ede87b} Edelson, R.~A.\ 1987, \apj, 313, 651
\bibitem[Edelson, Malkan, \& Rieke(1987)]{ede87} Edelson, R.~A., Malkan, 
M.~A., \& Rieke, G.~H.\ 1987, \apj, 321, 233 
\bibitem[Efstathiou \& Rowan-Robinson(1995)]{efs95} Efstathiou, A., \&
Rowan-Robinson, M.\ 1995, \mnras, 273, 649
\bibitem[Elvis et al.(2004)]{elv04} Elvis, M., Risaliti, G., Nicastro,
F., Miller, J.~M., Fiore, F., \& Puccetti, S.\ 2004, \apjl, 615, L25
\bibitem[Feigelson \& Nelson(1985)]{fei85} Feigelson, E.~D., \& Nelson, P.~I.\
1985, \apj, 293, 192
\bibitem[Francis \& Wills(1999)]{fra99} Francis, P.~J., \& Wills, B.~J.\ 1999,
ASP Conf.~Ser.~162: Quasars and Cosmology, 162, 363
\bibitem[Frank, King, \& Raine(1992)]{fra92} Frank, J., King, A.\ R., \&
Raine, D.\ J.\ 1992, Accretion power in astrophysics (2nd ed.; Cambridge: CUP)
\bibitem[Gallimore et al.(1999)]{gal99} Gallimore, J.~F., Baum, S.~A., O'Dea,
C.~P., Pedlar, A., \& Brinks, E.\ 1999, \apj, 524, 684
\bibitem[Gebhardt et al.(2000)]{geb00} Gebhardt, K., et al.\ 2000, \apjl, 539,
L13
\bibitem[Giuricin et al.(1990)]{giu90} Giuricin, G., Mardirossian, F.,
Mezzetti, M., \& Bertotti, G.\ 1990, \apjs, 72, 551
\bibitem[Gorjian et al.(2004)]{gor04} Gorjian, et al. 2004, ApJ, 605, 156
\bibitem[Granato \& Danese(1994)]{gra94} Granato, G.~L., \& Danese, L.\ 1994,
\mnras, 268, 235
\bibitem[Green, Schmidt \& Liebert(1987)]{gre87} Green, R.\ F., Schmidt, M.,
\& Liebert, J., 1987, \apjs, 61, 305
\bibitem[Hao et al.(2005a)]{hao05a} Hao, L., et al.\ 2005a, \aj, 
129, 1783 
\bibitem[Hao et al.(2005b)]{hao05b} Hao, L., et al.\ 2005b, \aj, 129, 1795
\bibitem[Hao et al.(2005c)]{hao05c} Hao, L., Spoon, H. W. W., Sloan, G. C.,
  Marshall, J. A., Armus, L., Tielens, A. G. G. M., Sargent, B., van Bemmel,
  I. M., Charmandaris, V., Weedman, D. W., Houck, J. R.\ 2005c,
  astro-ph/0504423
\bibitem[Heckman(1995)]{hec95a} Heckman, T.~M.\ 1995, \apj, 446, 101
\bibitem[Heckman et al.(1995)]{hec95b} Heckman, T., \etal\ 1995, \apj, 452,
549
\bibitem[Higdon et al.(2004)]{hig04} Higdon, S.\ J.\ U.\ et al.\ 2004, \pasp,
116, 975
\bibitem[Ho et al.(1997)]{ho97} Ho, L.~C., Filippenko, A.~V., \& Sargent,
W.~L.~W.\ 1997, \apjs, 112, 315
\bibitem[Houck et al.(2004)]{hou04} Houck, J.~R., et al.\ 2004, \apjs, 154, 18
\bibitem[Imanishi \& Alonso-Herrero(2004)]{ima04} Imanishi, M., \&
Alonso-Herrero, A.\ 2004, \apj, 614, 122
\bibitem[Ivezic \& Elitzur(1997)]{ive97} Ivezic, Z., \& Elitzur, M.\ 1997,
\mnras, 287, 799
\bibitem[Jaffe et al.(2004)]{jaf04} Jaffe, W., et al.\ 2004, \nat, 429, 47
\bibitem[Jogee et al.(2005)]{jog05} Jogee, S., Scoville, N., \& Kenney,
J.~D.~P.\ 2005, \apj, 630, 837
\bibitem[Kauffmann et al.(2003)]{kau03} Kauffmann, G., et al.\ 2003, \mnras,
346, 1055
\bibitem[Krolik \& Begelman(1988)]{kro88} Krolik, J.~H., \& Begelman, M.~C.\
1988, \apj, 329, 702
\bibitem[Kukula et al.(1995)]{kuk95} Kukula, M.~J., Pedlar, A., Baum, S.~A.,
\& O'Dea, C.~P.\ 1995, \mnras, 276, 1262
\bibitem[Lavalley et al.(1992)]{lav92} Lavalley, M., Isobe, T., \& Feigelson,
E.\ 1992, ASP Conf.~Ser.~ 25: Astronomical Data Analysis Software and Systems
I, 25, 245
\bibitem[Litchfield et al.(1989)]{lit89} Litchfield, S.~J., King,
A.~R., \& Brooker, J.~R.~E.\ 1989, \mnras, 237, 1067
\bibitem[Lutz et al.(1998)]{lut98} Lutz, D., Genzel, R., Kunze, D., Spoon,
H.~W.~W., Sturm, E., Sternberg, A., \& Moorwood, A.~F.~M.\ 1998, ASP
Conf.~Ser.~132: Star Formation with the Infrared Space Observatory, 132, 89
\bibitem[Lutz et al.(2004)]{lut04} Lutz, D., Maiolino, R., Spoon, H.~W.~W., \&
Moorwood, A.~F.~M.\ 2004, \aap, 418, 465
\bibitem[Maiolino \& Rieke(1995)]{mai95a} Maiolino, R., \& Rieke, G.~H.\ 1995,
\apj, 454, 95
\bibitem[Maiolino et al.(1995)]{mai95b} Maiolino, R., Ruiz, M., Rieke,
G.\ H., Keller, L.\ D.\ 1995, \apj, 446, 561
\bibitem[Mason et al.(2006)]{mas06} Mason, R.\ E., Geballe, T.\ R., Packham,
C., Levenson, N.\ A., Elitzur, M., Fisher, R.\ S., Perlman, E.\ 2006, \apj,
640, 2 (astro-ph/0512202)
\bibitem[Mauch et al.(2003)]{mau03} Mauch, T., Murphy, T., Buttery, H.~J.,
Curran, J., Hunstead, R.~W., Piestrzynski, B., Robertson, J.~G., \& Sadler,
E.~M.\ 2003, \mnras, 342, 1117
\bibitem[McAlary \& Rieke(1988)]{mca88} McAlary, C.~W., \& Rieke, G.~H.\ 1988,
\apj, 333, 1
\bibitem[Merritt \& Ferrarese(2001)]{mer01} Merritt, D., \& Ferrarese, L.\
2001, \apj, 547, 140
\bibitem[Nenkova et al.(2002)]{nen02} Nenkova, M., \& Ivezi{\' c}, {\v Z}., \&
  {Elitzur}, M.\ 2002, \apjl, 570, L9
\bibitem[Neugebauer \& Matthews(1999)]{neu99} Neugebauer, G., \& Matthews, K.\
1999, \aj, 118, 35
\bibitem[Norman \& Scoville(1988)]{nor88} Norman, C., \& Scoville, N.\ 1988,
\apj, 332, 124
\bibitem[Osterbrock(1993)]{ost93} Osterbrock, D.~E.\ 1993, \apj, 404, 551
\bibitem[Panagia \& Felli(1975)]{pan75} Panagia, N., \& Felli, M.\ 1975, \aap,
39, 1
\bibitem[Perez Garcia \& Rodriguez Espinosa(2001)]{per01} Perez
  Garcia, A.\ M., Rodriguez Espinosa, J.\ M.\ 2001, \apj, 557, 39
\bibitem[Phillips(1979)]{phi79} Phillips, M.~M.\ 1979, \apjl, 227, L121
\bibitem[Pier \& Krolik(1992)]{pie92} Pier, E.~A., \& Krolik, J.~H.\ 1992,
\apj, 401, 99
\bibitem[Pier \& Krolik(1993)]{pie93} Pier, E.~A., \& Krolik, J.~H.\ 1993,
\apj, 418, 673
\bibitem[Polletta et al.(2000)]{pol00} Polletta, M., Courvoisier, T.~J.-L.,
Hooper, E.~J., \& Wilkes, B.~J.\ 2000, \aap, 362, 75
\bibitem[Rees(1984)]{ree84} Rees, M.~J.\ 1984, \araa, 22, 471
\bibitem[Rieke(1978)]{rie78} Rieke, G.~H.\ 1978, \apj, 226, 550
\bibitem[Risaliti et al.(2005)]{ris05} Risaliti, G., Elvis, M.,
Fabbiano, G., Baldi, A., \& Zezas, A.\ 2005, \apjl, 623, L93
\bibitem[Rodriguez Espinosa et al.(1996)]{rod96} Rodriguez Espinosa,
J.\ M., Perez Garcia, A.\ M., Lemke, D., Meisenheimer, K.\ 1996, \aap,
315, L129
\bibitem[R{\" o}ttgering et al.(2005)]{rot05} R{\" o}ttgering, H., Jaffe, W.,
Meisenheimer, K., Sol, H., Leinert, C., Richichi, A., Wittkowski, M.\ 2005,
Proc.SPIE, 5491, "New Frontiers in Stellar Interferometry" (astro-ph/0507236)
\bibitem[Rowan-Robinson \& Crawford(1989)]{row89} Rowan-Robinson, M., \&
Crawford, J.\ 1989, \mnras, 238, 523
\bibitem[Rush et al.(1993)]{rus93} Rush, B., Malkan, M.~A., \& Spinoglio, L.\
1993, \apjs, 89, 1
\bibitem[Sanders et al.(1988)]{san88} Sanders, D.~B., Soifer, B.~T., Elias,
J.~H., Madore, B.~F., Matthews, K., Neugebauer, G., \& Scoville, N.~Z.\ 1988,
\apj, 325, 74
\bibitem[Shang \& Wills(2004)]{sha04} Shang, Z., \& Wills, B.\ 2004, ASP
Conf.~Ser.~311: AGN Physics with the Sloan Digital Sky Survey, 311, 13
\bibitem[Shields(1999)]{shi99} Shields, G.~A.\ 1999, \pasp, 111, 661
\bibitem[Siebenmorgen et al.(2005)]{sie05} Siebenmorgen, R., Haas, M.,
  Kr{\" u}gel, E., Schulz, B.\ 2005, astro-ph/0504263
\bibitem[Siebenmorgen et al.(2004)]{sie04} Siebenmorgen, R., Kr{\" u}gel, E.,
\& Spoon, H.~W.~W.\ 2004, \aap, 414, 123
\bibitem[Spinoglio \& Malkan(1989)]{sm89} Spinoglio, L., \& Malkan,
  M. A. 1989, \apj, 342, 83
\bibitem[Storchi-Bergmann et al.(2001)]{sto01} Storchi-Bergmann, T.,
Gonz{\'a}lez Delgado, R.~M., Schmitt, H.~R., Cid Fernandes, R., \& Heckman,
T.\ 2001, \apj, 559, 147
\bibitem[Thean et al.(2000)]{the00} Thean, A., Pedlar, A., Kukula, M.~J.,
Baum, S.~A., \& O'Dea, C.~P.\ 2000, \mnras, 314, 573
\bibitem[Thean et al.(2001)]{the01} Thean, A., Pedlar, A., Kukula, M.~J.,
Baum, S.~A., \& O'Dea, C.~P.\ 2001, \mnras, 325, 737
\bibitem[Tran(2003)]{tra03} Tran, H.~D.\ 2003, \apj, 583, 632
\bibitem[Urry \& Padovani(1995)]{urr95} Urry, C.~M., \& Padovani, P.\ 1995,
\pasp, 107, 803
\bibitem[Veilleux et al.(1995)]{vei95} Veilleux, S., Kim, D.-C., Sanders,
D.~B., Mazzarella, J.~M., \& Soifer, B.~T.\ 1995, \apjs, 98, 171
\bibitem[Ulvestad et al.(1999)]{ulv99} Ulvestad, J.~S., Wrobel, J.~M., Roy,
A.~L., Wilson, A.~S., Falcke, H., \& Krichbaum, T.~P.\ 1999, \apjl, 517, L81
\bibitem[V{\' e}ron-Cetty \& V{\' e}ron(2003)]{ver03} V{\' e}ron-Cetty, M.-P.,
\& V{\' e}ron, P.\ 2003, \aap, 412, 399
\bibitem[Ward et al.(1987)]{war87} Ward, M., Elvis, M., Fabbiano, G.,
Carleton, N.~P., Willner, S.~P., \& Lawrence, A.\ 1987, \apj, 315, 74
\bibitem[Weedman et al.(2005)]{wee05}  Weedman, D. W., Hao, L., Higdon,
  S. J. U., Devost, D.,  Wu, Y., Charmandaris, V., Brandl, B., Bass, E.,
  Houck, J. R.\ 2005, astro-ph/0507423
\bibitem[White et al.(1997)]{whi97} White, R.~L., Becker, R.~H., Helfand,
D.~J., \& Gregg, M.~D.\ 1997, \apj, 475, 479
\bibitem[Xu et al.(1999)]{xu99} Xu, C., Livio, M., \& Baum, S.\ 1999, \aj,
118, 1169
\end{thebibliography}
\end{document}